\documentclass[a4paper]{aa}
\usepackage{xspace}
\usepackage{graphicx}
\usepackage{color}
\usepackage{txfonts}
\usepackage{natbib}
\usepackage{lscape}
\usepackage{longtable}
\usepackage{balance}
\usepackage{placeins}
\usepackage{paralist}
\usepackage{xtab,booktabs}
\usepackage{multirow}
\usepackage[breaklinks=true,bookmarks=true,hypertexnames=false,bookmarksnumbered=true,bookmarksopen=false,bookmarks=true,bookmarksnumbered=true]{hyperref}
\usepackage[]{color}

\definecolor{Red}{rgb}{0.65,0.08,0.05}

\newcommand{\msun}{{\rm M}_{\odot}}

\newcommand{\h}{\,h}
\newcommand{\hmm}{\h^{-1}}
\newcommand{\hmmsun}{\hmm\msun}

\newcommand{\der}{{\rm d}}

\newcommand*\Email[1]{E-mail: {\aa@emailfont #1}}

\def\rmax{ R_{\rm max} }
\def\rmin{ R_{\rm min} }

\newcommand\T{\rule{0pt}{2.6ex}}       % Top strut
\newcommand\B{\rule[-1.2ex]{0pt}{0pt}} % Bottom strut

\begin{document}

\title{Multipolar moments of weak lensing signal around clusters}
\subtitle{Weighing filaments in harmonic space}

\author{
  C.~Gouin\inst{1}\thanks{E-mail:~\tt{gouin@iap.fr}},
  R.~Gavazzi\inst{1},
  S.~Codis\inst{2},
  C.~Pichon\inst{1,3},
  S.~Peirani\inst{1},
  Y.~Dubois\inst{1}
}

\institute{
Institut d'Astrophysique de Paris, UMR7095 CNRS \& Universit\'e Pierre et Marie Curie, 98bis Bd Arago, F-75014, Paris, France
\label{inst1}
\and
Canadian Institute for Theoretical Astrophysics, University of Toronto, 60 St. George Street, Toronto, ON M5S 3H8, Canada
\label{inst2}
\and
Korea Institute of Advanced Studies (KIAS) 85 Hoegiro, Dongdaemun-gu, Seoul, 02455, Republic of Korea
\label{inst3}
}

\date{\today}

%------------------------------------------------------------------------------ 

\abstract
%context
{
Upcoming weak lensing surveys such as Euclid will provide an unprecedented opportunity to quantify 
the geometry and topology of the cosmic web, in particular in the vicinity of lensing clusters. 
}
%aim
{ 
Understanding the connectivity of the cosmic web with unbiased mass tracers, such as weak lensing, is of prime importance to probe the underlying cosmology, seek dynamical signatures of dark matter, and quantify  environmental effects on galaxy formation.
}
%method
{ 
Mock catalogues of galaxy clusters are extracted from the N-body PLUS simulation.
For each cluster,  the aperture multipolar moments of the convergence are calculated in two annuli (inside and outside the virial radius). 
By stacking their modulus, a statistical estimator is built to characterise the angular mass distribution around clusters.
The  moments are compared  to predictions from perturbation theory and spherical collapse.
}
%results 
{
The main weakly chromatic excess of multipolar power on large scales is understood as arising from the contraction of the primordial cosmic web driven by the growing potential well of the cluster.
Besides this boost, the quadrupole prevails in the cluster  (ellipsoidal) core, while at the outskirts, harmonic distortions are spread on small angular modes, and  trace the non-linear sharpening of the filamentary structures. 
Predictions for the signal amplitude as a function of the cluster-centric distance, mass, and redshift are presented.  
The prospects of measuring this signal are estimated for current and future lensing data sets.
}
%conclusion
{
The Euclid mission should provide all the necessary information for studying the cosmic evolution of the connectivity of the cosmic web around lensing clusters using multipolar moments and probing unique signatures of, for example, baryons and warm dark matter.
}

   
\authorrunning{Gouin et al.}
\titlerunning{Weighing filaments from multipolar moments of the convergence}

\maketitle

%------------------------------------------------------------------------------ 
%------------------------------------------------------------------------------ 
\section{Introduction}
The large-scale structures of the Universe (hereafter LSS) have been observed for more than twenty years, first by the CfA catalogue \citep{DeLapparent+1986}, and later by large spectroscopic surveys (e.g. the 2dF and SDSS surveys : \citealt{Colless+03, Gott+05}). These surveys emphasised a distribution of galaxies that is not homogeneous, but concentrated along filaments, sheets, and presents large underdense regions (voids). Structures grow highly anisotropically from primordial density fluctuations of dark matter under the effect of gravity. Large N-body simulations reproduce this picture, where dark matter particles arrange themselves in a network of filaments. Massive haloes are located at the intersection of filaments, and grow via successive merging as well as continuous accretion of surrounding matter following some preferential directions \citep{davis85,ks93,Bond+96,bm96a}.

The detection and study of cosmic filaments through observations and numerical simulations is a fundamental step in refining our understanding of structure formation history and cluster evolution. Indeed, filaments have a significant impact on the mass budget of the Universe, as they account for $\sim 40$ per cent of all mass at $z=0$ \citep{Aragon+10}.  Their dynamical evolution probes the underlying cosmological model.  They also play an important environmental role on galaxy formation and galaxy properties \citep{hahn07,Sousbie+08,Pichon+11,Codis+12,Malavasi++17,Laigle++17}.  However, given the low-density contrast of filaments, identifying them in observations has remained a challenge and, in this context, several observables have been devised  to probe their  mass distribution.
 
One of the main observables that has been used to detect filaments is X-ray emission, induced by the warm hot intergalactic medium  \citep[WHIM,][]{Cen+99}. Detection remains difficult because X-rays could either come from the intergalactic medium, or past cluster mergers \citep[e.g.][]{Kull+99, Durret+03}, and often needs to be supported by other observables \citep{Eckert+15}. Recently, the Planck satellite has also claimed detection of the Sunyaev-Zeldovich effect of the WHIM in between pairs of galaxy clusters \citep{planck8}.
Some studies reported the detection of filaments in the distribution of galaxies \citep{Ebeling+04, Pimbblet+04}. Yet this method is limited to relatively low redshifts ($z<0.4$) and does not probe the physical properties of filaments, given that its main components are the WHIM and dark matter. \cite{Zhang+13} proposed an algorithm to study the photometric properties of filaments  by stacking galaxies' population in cluster pairs.

Gravitational lensing stands  as a powerful complementary  tool to investigate the entire structure of filaments because it probes dark and luminous matter, regardless of its dynamical state. However,  early filament detections by weak lensing (WL) are still controversial. Some attempts have been made to detect them in between galaxy clusters \citep{ Clowe+98, K98, Gray+02, Gavazzi+04, Dietrich+04,Heymans+08}, but it is not clear how robust these results are against residual systematic galaxy alignments.
Progress in mass reconstructions from WL (alone or in combination with strong lensing)  have  recently led to  detections   confirmed by other observables \citep{Dietrich+12, Jauzac+12, Higuchi+15}. Yet,  direct measurements from individual  (massive) filaments remain sparse due to the weakness of the corresponding shear field. To compensate, some studies also tried to characterise filaments by stacking WL signal \cite[see e.g. ][]{Dietrich+05, Mead+10, Clampitt+16, Epps+17}.
Various WL algorithms  amplifying  filament detection are described in the literature. \citet{Maturi+13}  developed a WL filter tailored to their elongated extension. \citet{Simon+08} used higher-order correlations, like the galaxy-galaxy-shear three-point statistics.
Likewise, aperture multipole moments were introduced by \citet{SB97} to quantify asymmetries in the mass distribution directly from the shear signal carried by background galaxies. \citet{Dietrich+05} and \citet{Mead+10} studied  the quadratic aperture moment of WL signal induced by cluster pairs.

Beyond the overall radial mass profile, the WL measurement of the ellipticity of haloes (their quadrupole) has drawn some attention in recent years. In particular, for galaxy  haloes, it is possible to study the relative alignment and flattening of the host halo and its central galaxy \citep{parker07,vanUitert12,Schrabback15,Clampitt16b}. 
Some attempts have also been made to measure the projected ellipticity of a few individual clusters of galaxies \citep{Ogu++10} or  to measure the mean ellipticity of an ensemble of clusters and groups of galaxies by stacking the signal \citep{evans09,vanUitert16}. This latter approach requires assumptions to be made over the relative alignment and elongation between dark and luminous matter.

 A visual inspection of the time evolution of N-body simulations allows us to anticipate the following: as gravitational clustering builds up,
to first order, the cosmic nodes catastrophically  attract matter in their vicinity.  This induces an amplification of the contrast in the 
connected cosmic network, which is locally  evolving more rapidly  due to the  induced density boost (when compared  to typical filaments away  from the nodes). At the level of  this spherically contracting description  there should be an excess harmonic power near the peaks.
At second order, the filaments themselves  induce anisotropic tides  which  boost up their own contrast by  transversally collecting  matter 
and substructures. This effect is also reinforced near peaks, as the radial and transverse tides add up
and proto-haloes pass the collapse threshold more easily. The  local cosmic  filaments at the nodes are therefore  amplified by non-linear gravity. 
This is the cluster-centric counterpart  of  
the  process described by \cite{Bond+96} for the field: the large-scale  cosmic web is de facto already in place in the initial conditions and
gravitational clustering amplifies it differentially. A relative harmonic analysis of the vicinity of clusters should therefore allow us to
 capture this gravitationally  boosted primordial connectivity.
 The purpose of this paper is to quantify this effect via upcoming WL surveys.

This study will rely on aperture multipole moments at all orders to quantify the azimuthal repartition of matter at different scales, centred on galaxy clusters. We will use  mock clusters extracted from a large N-body cosmological simulation to predict the statistical properties of  multipolar moments.  A new statistical estimator, the multipolar power spectrum will be implemented while stacking the modulus of aperture multipole moments. 
This method will allow us to quantify the angular distribution of matter around cosmic hubs, hence to detect the signature of filaments in the vicinity of clusters. Stacking power spectra instead of moments alleviates  assumptions about the relative distribution of dark and luminous matter whose relation seems to depend on mass and scale \citep{vanUitert16}.

The structure of this paper is as follows.
Sect.~\ref{sec:anal} describes the aperture multipole moments following the formalism of \citet{SB97} and relates the statistics of these moments with those of the underlying convergence fields. In this section, we also build a model for the expected boost in the harmonic power spectra.
Sect.~\ref{sec:simus}  then describes the computation of multipole moments using the dark halo clusters extracted from the PLUS cosmological constrained dark matter N-body simulation and  explores how the power spectra depend on redshift, cluster mass and radius.
 Sect.~\ref{sec:snr}  discusses the prospects of measuring this signal with WL data accounting for sample variance, shape noise (finite ellipticity of background sources), and intervening LSS, and then weighs the mass content of filaments near the nodes of the cosmic web. 
 Finally, a summary is presented in Sect.~\ref{sec:conclusion}.

%------------------------------------------------------------------------------ 
%------------------------------------------------------------------------------ 
\section{Multipolar aperture moments}\label{sec:anal}

Let us first present the  ingredients  of cluster-centric weak gravitational lensing and introduce  the corresponding expected statistical properties of the convergence  field, which is related to  the underlying matter distribution. Specifically,  multipolar moments  are introduced to measure the asphericity -- which quantifies the projected departures from circular symmetry -- around the nodes of the cosmic web. Since the focus is on  how cluster environment deviates from random locations,  the  expected ratios will be presented in  increasing order of theoretical  complexity,    starting from the assumption of  constrained Gaussian random fields (GRFs) for the convergence. This will guide our understanding of the actual empirical numerical study discussed in the following section.

%.......................................
\subsection{Definition of convergence multipoles }\label{ssec:def}
The focus of this paper lies in the azimuthal mass distribution at various scales
around massive galaxy clusters.
 For a thin gravitational lens plane,
the convergence $\kappa$  at a given
position $\vec{r}$ in the sky corresponds to the projected excess surface density expressed
in units of the so-called critical density $\Sigma_{\rm crit}$
\begin{equation}
\kappa(\vec{r}) = \frac{1}{\Sigma_{\rm crit}} \int \der z
\left( \rho(\vec{r},z) - \overline{\rho} \right) \,,
\end{equation}
with the convention that the line-of-sight corresponds to the $z$-axis
and the plane of the sky $\vec{r}$ vector can be defined by polar
coordinates $(r,\varphi)$. The critical density involves distance
ratios between a fiducial source at an angular diameter distance
$D_{\rm s}$, the distance to the lensing mass $D_{\rm l}$ and the
distance between the lens and the source $D_{\rm ls}$
\begin{equation}\label{eq:scrit}
  \Sigma_{\rm crit} = \frac{c^2}{4 \pi G} \frac{ D_{\rm s}}{ D_{\rm l}
    D_{\rm ls}}\,.
\end{equation}
The lensing potential $\psi$ can also be defined as
\begin{equation}
\psi(\vec{r}) = \frac{2}{c^2}\frac{D_{\rm l}D_{\rm ls}}{ D_{\rm s}} \int\der z \,\Phi(\vec{r},z) \,,
\end{equation}
where $\Phi$ is the three dimensional gravitational potential.  One can then express the two components $\gamma_1$ and $\gamma_2$  of the complex spin-2 shear $\gamma=\gamma_1+ \imath \gamma_2$ and the scalar convergence $\kappa$ as derivatives of the lensing potential
\begin{eqnarray}
2 \kappa  & = & \psi_{,11} + \psi_{,22}\,, \\
2 \gamma_1 &=& \psi_{,11} - \psi_{,22}\,, \\
\gamma_2 &=& \psi_{,12}\,.
\end{eqnarray}
Since the interest lies in convergence and shear patterns around a particular centre, let us introduce the tangential $\gamma_t $ and curl $\gamma_\times$ shear components\footnote{In some cases, $\gamma_\times$ is referred to as a ``radial''  $\gamma_r$ component which can be misleading for a spin-2 field.} as
\begin{equation}
        \gamma_t   = - \mathcal{R}(  \gamma \, {\rm e}^{-2 \imath  \varphi})\,, \quad
        \gamma_\times =  - \mathcal{I}(  \gamma \, {\rm e}^{-2 \imath  \varphi})\,.
\end{equation}
Following the formalism of \citet{SB97}, let us also define the aperture multipole
moments $Q_m$ of this convergence field as
\begin{equation}\label{eq:mpoledef}
  Q_m = \int_0^\infty \der r\,  r^{1+m} w_m(r) \int_0^{2\pi} \der
  \varphi\,  {\rm e}^{\imath  m \varphi} \kappa(r,\varphi) \,.
\end{equation}
There is substantial margin in the choice of the radial weight function. As one may want to have a different radial shape for different multipole orders $m$ and it is desirable to consider a compact support so that $w_m(r)$ vanishes beyond some radius $\rmax$ and, possibly, below some inner radius $\rmin \equiv \nu \rmax$.  For later use,   the multipolar moments can be expressed as a function of the shear field \citep{SB97} in a local
\begin{eqnarray}\label{eq:mpoledef2l}
  Q_m & = & \int_0^\infty \der r\,  r^{1+m}  \int_0^{2\pi} \der \varphi\,  {\rm e}^{\imath  m \varphi}\times  \nonumber\\
    &&  \left[ w_m(r)  \gamma_t(r,\varphi)   +  \imath  \left( w_m(r) + \frac{r}{m} w_m'(r)\right) \gamma_\times(r,\varphi) \right]  \,,
\end{eqnarray}
and a non-local 
\begin{eqnarray}\label{eq:mpoledef2nl}
  Q_m & = & \int_0^\infty \der r\,  r^{1+m}  \int_0^{2\pi} \der \varphi\,  {\rm e}^{\imath  m \varphi} \times \nonumber\\
    &&  \left[ \left(2 W_m(r)-w_m(r)\right)  \gamma_t(r,\varphi)   -  \imath  m W_m(r) \gamma_\times(r,\varphi) \right]  \,,
\end{eqnarray}
way,  with the long-range weight function defined by 
\begin{equation}
     W_m(r)  = \frac{1}{r^{m+2}}\int_0^r \der x\, x^{1+ m} w_m(x)\,,
\end{equation}
where the prime denotes $\der\, /\der r$ derivation.
Any combination of these two estimators would yield the same answer but, in practice, a careful account of the various sources of noise and the range over which data (i.e. ellipticities of background galaxies) are available will drive the choice of $w_m(r)$. By considering the main source of noise that is due to the intrinsic non-zero ellipticity distribution of background galaxies, \citet{SB97} found an optimal weight function for a mass density profile that can be approximated  as a nearly isothermal mass distribution with $\rho(r)\propto r^{-2}$ or ($\kappa\propto r^{-1}$) and proposed the following form that we use, unless otherwise stated:
\begin{equation}\label{eq:weightSB7}
\rmax^{1+m} \, w_m(r)  = \frac{1}{x^{1+m}+\nu^{1+m}} - \frac{1}{1+\nu^{1+m}}   + \frac{(1+m)(x-1)}{(1+\nu^{1+m})^2}\,,
\end{equation}
over the range $x=r/\rmax \in [\nu, 1]$ and zero elsewhere. 

It is noteworthy that an elliptical mass distribution will generate only even moments with a fast decline of modes. \citet{S+W91} relate the harmonic expansion terms of the convergence or surface mass density with the ellipticity $\epsilon$. In practice, for a power law mass distribution $\kappa \propto r^{-n}$, one can easily show that $Q_2/Q_0 = n \epsilon / 2$.

%.......................................
\subsection{The statistics of $Q_m$ for a Gaussian random field}\label{ssec:anal2pt}

In this subsection,  let us further assume that the convergence (or some projected
density) field is Gaussian and fully characterised by its power
spectrum $P_\kappa(\vec{k})$ so that the two-point expectation value for the
Fourier modes $\hat{\kappa}(\vec{k})$ can be written as
\begin{equation}
  \langle \hat{\kappa}(\vec{k}) \hat{\kappa}^*(\vec{k'}) \rangle = (2
  \pi)^2 \delta_D(\vec{k}+\vec{k'}) P_\kappa(\vert \vec{k}\vert) \,.
\end{equation}

%................
\subsubsection{Random locations}\label{ssec:anal2ptRL}
The convergence $\kappa$ having zero mean value,  the covariance between multipolar moments centred at random positions can readily
be expressed by following the same method as \citet{schneider98} who explored the statistics of the
$M_{\rm ap}$ statistic, which is a particular case of the $m=0$
multipolar moments. Let us write
\begin{eqnarray}\label{eq:mpolecov}
  \langle Q_m Q_n^* \rangle  & =  &
 \iint\limits_0^\infty r\der r\, r'\der r' \,  \iint\limits_0^{2\pi} \der
  \varphi\der  \varphi'\, \,  r^{m} w_m(r)  r'^{n} w_n(r')  \nonumber\\
&& \;\;\times\; {\rm e}^{\imath  (m \varphi - n\varphi')} \langle
  \kappa(r,\varphi) \kappa(r',\varphi') \rangle\,,\\
&= & 2 \pi \, \imath^{m-n} \int k\der k\, U_m(k) U_n(k)\, P_\kappa(k) \,,\label{eq:mpolecovsub}\\
&\equiv & \imath^{m-n} A_{\kappa,mn} \,,
\end{eqnarray}
where    the Hankel transform $U_n(\ell)$  of the radial weight function is defined by 
\begin{equation}\label{eq:weightHankel}
   U_m(\ell) = \int r \der r\, r^m w_m(r)  J_m(\ell r)\,,
\end{equation}
and $J_m(x)$ are the first-kind Bessel functions. 

\begin{figure}
\includegraphics[width=0.47\textwidth]{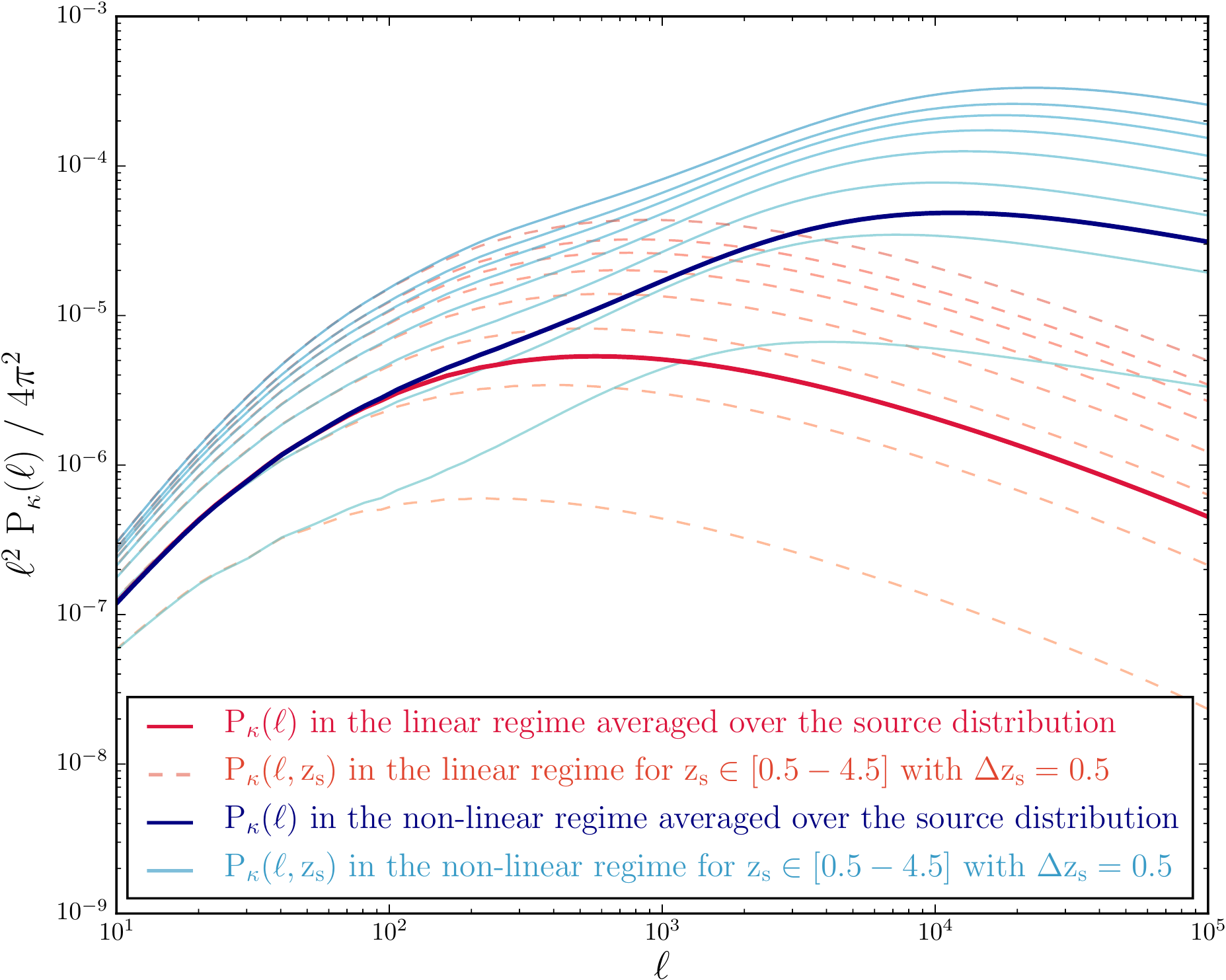}  
\includegraphics[width=0.47\textwidth]{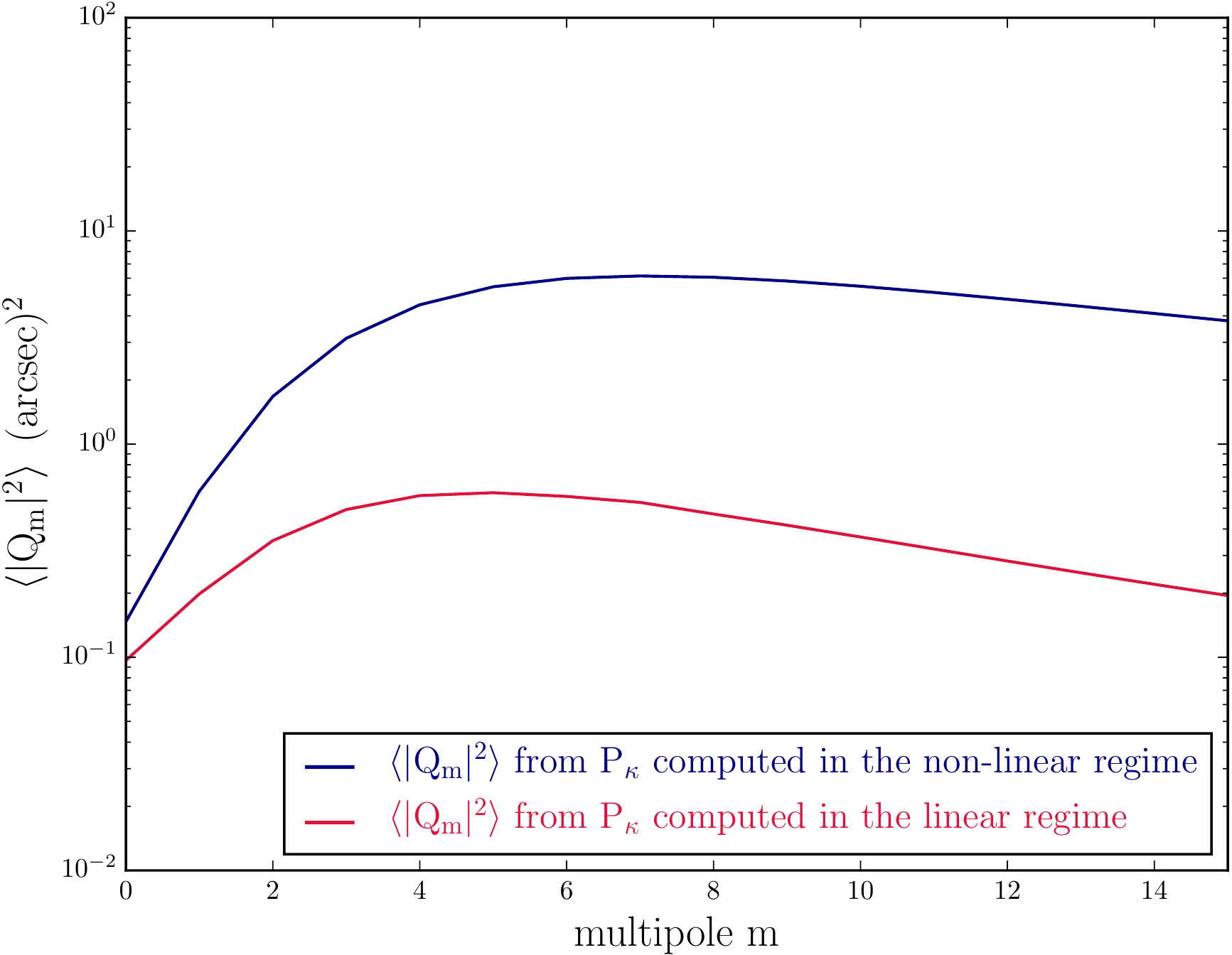}  
\caption{{\it Top panel:} Cosmological convergence power spectrum $P_\kappa(\ell)$ derived from the evolved matter power spectrum averaged over the source redshift distribution of Eq.~\eqref{eq:nofz}. The red and blue curves represent the quantities induced by the linear and non-linear matter power spectrums,  respectively.\ \ \ \ \ {\it Bottom panel:} Corresponding multipolar moment spectra $\langle \vert Q_m^2 \vert \rangle$ for the weight function in Eq.~\eqref{eq:weightSB7} and a choice of $R_{\rm max}=10\arcmin$ and $\nu=0.5$. }
\label{fig:qncosmo}
\end{figure}

Let us now consider a realistic convergence power spectrum   derived from the non-linearly evolved matter  spectrum, $P_\delta(\vec{k})$.   The power spectrum of the density contrast is computed at various redshifts,  using the Boltzmann code {\tt CLASS} toolkit \citep{class,class2} for the fiducial Planck Cosmology. For a given source redshift, the convergence power spectrum $P_\kappa(\ell,z_s)$ can be inferred from the three-dimensional matter power spectrum, considering the following integral \citep{blandford91,miralda91,K92,BS01,Simon+07}
\begin{equation}
P_\kappa(\ell,z_s) = \frac{9}{4} \Omega_m^2 \left( \frac{H_0}{c} \right)^4 \int_0^{\chi_{s}} \der \chi \frac{(\chi_s - \chi)^2}{\chi_s^2 } \frac{P_{\delta}\left({\ell}/{\chi},\chi \right)}{a^2(\chi)} \;.
\label{eq:pk_ell}
\end{equation}
In addition, instead of using a single-source-plane redshift,  the latest COSMOS2015 photometric redshift distribution \citep{Laigle16} is used to approximate the redshift distribution of sources as a Gamma PDF of the form
\begin{equation}\label{eq:nofz}
p(z_{\rm s}) = \frac{{\rm e}^{-z_{\rm s}/z_0}}{z_0 \Gamma(a)} \left(\frac{z_{\rm s}}{z_0}\right)^{a-1}\;,
\end{equation}
with $a\simeq 2.1$ and $z_0 \simeq 0.51$ for sources as faint as an AB magnitude $i=25,$ which is suitable for future experiments like Euclid or current deep ground-based imaging data.
The effective convergence power spectrum is finally computed by weighting the contributions of the different source planes
\begin{equation}
P_\kappa(\ell) = \int P_\kappa(\ell,z_s) \ p(z_{\rm s}) \ \der z_s \;,
\end{equation}
For illustration purposes,  Fig.~\ref{fig:qncosmo} shows the convergence power spectrum and the corresponding  multipolar moments for a fiducial choice of $R_{\rm max}=10\arcmin$ and $\nu=0.5,$ which should correspond to the scale of a few virial radii for a massive cluster at redshift $z\sim 0.3$. In order to highlight the contrast between the linear and non-linear regime on the convergence power spectrum and on the associated multipolar moment spectrum, we illustrate these two cases. As expected, by adopting a non-linear
matter power spectrum, the multipolar moment spectrum is significantly enhanced.

%%%%.........

%................
\subsubsection{The statistics of $Q_m$ under peak constraint}\label{ssec:th3pt}
Even for a Gaussian random field, the statistics of $Q_m$ should change
significantly  when centred on a cluster rather than a random location. The formalism must be updated  to deal
with a particular flavour of three-point statistics that accounts for the
presence of a maximum with a specific height $\nu_{p}$ of the density field at the origin of the
coordinate system. 
In brief, the companion paper \citep{Codis++17} has shown that the effect of the peak constraint is to 
\begin{itemize}
\item significantly boost the monopole (we are near a peak),
\item significantly remove power from the dipole (we are now well centred on the peak),
\item slightly suppress the power of the quadrupole,
\item leave all other $m\ge3$ multipoles unchanged.
\end{itemize}
These results rely on the assumptions that galaxy clusters can be mapped to peaks in the initial field smoothed at some scale $R$, which themselves can be characterised by their height (large excursion), gradient (forced to be null at the origin), and Hessian (two negative eigenvalues).

Our current purpose is to go beyond the Gaussian and peak approximation and describe the statistical properties of $Q_m$ at late time by measuring them directly in simulations in Sect.~\ref{sec:sim}.
In the following section, we simply describe changes in the statistics of multipolar moments to be expected from simple arguments about the
 nonlinear evolution around galaxy clusters.

%.......................................
\subsection{The nonlinear statistics of $Q_m$ around clusters}\label{ssec:analZel}
Appendix~\ref{app:zeldo} presents an approximate model based on the Zeldovich approximation and the spherical collapse, which shows  that the small-scale density fluctuations in a shell at radius $r$ falling onto a spherically symmetric proto-cluster will experience a boost of amplitude $\alpha$ with respect to the field.
It is due to the contraction of fluctuations within the original Lagrangian shell induced by  the tidal distortion field of the cluster overdensity.
It implies for the 3D power spectrum 
\begin{equation} 
P_{\rm cluster} (\vec{k})  \equiv  \alpha\; P_{\rm random}(\vec{k})\,, \quad {\rm where }\quad 
\label{eq:alpha0}
\alpha \simeq \left( \frac{3 M(<r) }{4\pi \bar{\rho} r^3} \right)^{2/3}, 
\end{equation}
where $\bar \rho$ is the mean background density. Once projected along the line-of-sight, fluctuations around a cluster
should also produce a nonlinear achromatic ($m$-independent) boost of multipolar moments spectra as compared to the field.
This prediction is compared to measurements in simulations in Sect.~\ref{resultz0} and
despite a crude treatment of projection effects and the extension of the model deeply inside the core of the cluster, is shown to give a quantitative explanation for the boost.

This boost corresponds to the first order change on the multipolar moments expected for initial peaks evolving into clusters. 
The next step involves understanding any spectral distortion of $\vert Q_m\vert^2_{\rm cluster} $ with respect to $ \alpha \vert Q_m\vert^2_{\rm random}$, as a second-order effect due to the  nonlinear coupling of modes involving specifically the filamentary structure around clusters.
Finally, we recall that the linear power spectrum $P(k)$ should be updated when accounting for the peak condition at the origin of the coordinate system and that, in two dimensions, the multipoles $m\le 2$ are impacted. 
 A more thorough treatment of the asphericity around the cluster potential could possibly allow us to predict the actual shape of the $\vert Q_m\vert^2_{\rm cluster} $ spectrum at all $m$. This is left for future work, and from now on, we rely on N-body cosmological simulations to predict these spectra, both near the edge of clusters where our formalism should hold, but also deep inside the virial radius of those clusters.

%------------------------------------------------------------------------------ 
%------------------------------------------------------------------------------ 
\section{Measuring multipolar moments in  simulations}\label{sec:simus}
\label{sec:sim}
%.......................................
\subsection{Dark matter haloes in the PLUS cosmological simulation}\label{ssec:plus}
\label{sec:simPlus}

%%%%.........
\begin{figure}
\includegraphics[width=0.49\textwidth]{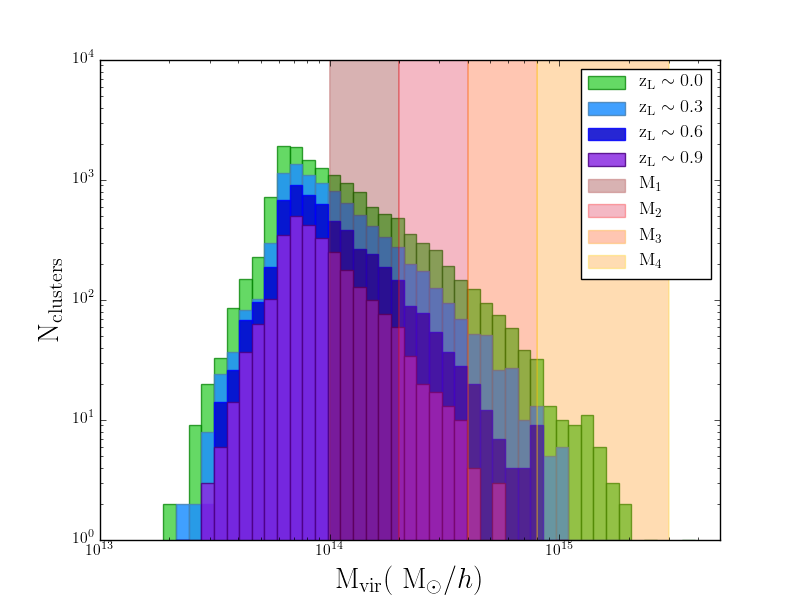}  
\caption{Halo virial mass function in the PLUS simulation box ($600\, {\rm cMpc}/h$ on a side) at four different redshifts (0, 0.3, 0.6 and 0.9) showing the content of the four mass intervals  considered. Only haloes with a FOF mass greater than $5\times10^{13}\hmmsun$ are shown here.}
\label{fig:histomvir}
\end{figure}

 A cosmological simulation taken from the
Paris Local Universe Simulation (PLUS)\footnote{\url{http://www2.iap.fr/users/peirani/PLUS/plus.htm}}
project was analysed. It corresponds to  a   $\Lambda$CDM universe  with the following set of cosmological parameters,
$\Omega_m=0.3175$,  $\Omega_{\Lambda}=0.6825$, $\Omega_b=0.049$, 
$H_0=67.11$ km/s/Mpc, $n_s=0.9624$ and $\sigma_8=0.8344$ \citep{Planck2013xvi}.
 The simulation was performed with Gadget2 \citep{Springel05}
in a periodic box of side $600\,h^{-1}$ Mpc and using $2048^3$ dark matter particles (i.e., with a mass resolution
of $\sim 2.2 \times 10^9\,h^{-1}M_\odot$). The adopted Plummer-equivalent force softening was 14.6  $h^{-1}$ kpc
and was kept constant in comoving units. The simulation  started at $z=49$ and ended at the present time $z=0$.
The initial conditions have been generated using the BORG algorithm 
\citep[Bayesian Origin Reconstruction from Galaxies][]{J+W13,lavaux15} 
aiming at modelling the local universe.
 
The   dark matter  halo catalogue was extracted at redshifts $z=0, 0.3, 0.6$, and $0.9$ with a Friend-of-Friend algorithm (FOF) using a linking length of 0.15 in
units of the mean inter-particle separation and came up, for the $z=0$ output, with about $14\,000$ groups and clusters with mass $M_{\rm FOF}>5\times10^{13} \hmmsun$. Then  all the particles in the direct vicinity of these haloes are extracted and projected along a given direction. Hence,  each extracted halo contains a cluster and its outer environment \citep{metzler01}. The additional effect of the uncorrelated background and foreground matter distribution along the line of sight to distant gravitationally lensed galaxies will be treated as an additional Gaussian random field acting as a noise contribution \citep[see e.g.][]{hoekstra01b,Hoekstra+03}.

\begin{itemize}
\item $M_{\rm FOF}$ is used as a first guess to define a virial radius $M_{\rm FOF} = 4/3 \pi \Delta_{\rm vir} \rho_{\rm crit} R_{\rm vir}^3$.
\item The centre of mass of the linked particles was used to extract all the particles within a comoving radius $R_{\rm H} = 4 R_{\rm vir}$ about it. This radius is sufficiently large to capture the most relevant environment of clusters.
\item We refine the definition of the centre by seeking the main peak of the density field with a shrinking sphere method. Starting from the previous value of $ R_{\rm vir}$, let us compute the centre of mass therein. At each iteration, the sphere is shrunk by 2.5\% and we update the center of mass accordingly. The process is stopped when the final mass is below 1\% of the starting $M_{\rm FOF}$ value.
\item By sorting particles in radius about this final centre, one can easily build the cumulative mean density profile
 $\overline{\rho}(<r)$. The final virial radius is the distance at which $\overline{\rho} (<R_{\rm vir}) = \Delta_{\rm vir} \rho_{\rm crit} $.
\end{itemize}
The above calculations rely on the fitting functions of \citet{Bryan1998} to estimate the density contrast $\Delta_{\rm vir}$ above the critical density $\rho_{\rm crit}$ for our reference $\Lambda$CDM cosmology. Hence, the typical virial mass is on average 1.14 times greater than  $M_{\rm FOF}$,  the mass directly linked by the FOF algorithm.
We investigate two different radial intervals, $R/r_{\rm vir} \in [0.25, 0.5] $ and $R/  r_{\rm vir} \in [1, 4]$, in order to emphasise differences between the innermost, presumably relaxed, areas and the ones undergoing coherent infall motions, where filaments should be more prominent.

To study the influence of mass in this analysis, the sample of haloes is divided into four bins  of virial mass  $M_1  \in 1-2 $, $M_2 \in 2-4$, $M_3 \in 4-8$, and $M_4 \ge 8 \times 10^{14}\hmmsun $ as shown in Fig.~\ref{fig:histomvir}.

We also study the evolution of multipolar moments with redshift by considering simulation outputs at redshifts $z=0,0.3,0.6,0.9$. 
In order to follow the evolution of moments of a given population of haloes, we also consider the 100 most massive clusters.  This selection in mass  is a proxy for a population of haloes with the same rareness, $\nu_{p}=\delta/\sigma$. Indeed, the mass of non-linearity $M_{\star}$ (i.e. the peak of the Press-Schechter mass function) evolves with redshift. Picking the most massive haloes at each redshift therefore allows us to focus on a population of haloes which present a similar level of non-linearity\footnote{which amounts  to having a constant variance $\sigma^2(M,z)$ at redshift $z$ and scale $M$.}, which should therefore be at the same stage of their evolution.  As redshift takes values $z=0, 0.3, 0.6,$ and $0.9$, the mean virial mass of the 100 most massive clusters successively takes values  $M_{\rm vir} \simeq  10,\, 7,\,  5,\,  3\, \times 10^{14} \hmmsun$.

%.......................................
\subsection{Multipoles from simulated haloes}

%%%%.........
\begin{figure*}%[!h]
\centering
\includegraphics[width=0.95\textwidth]{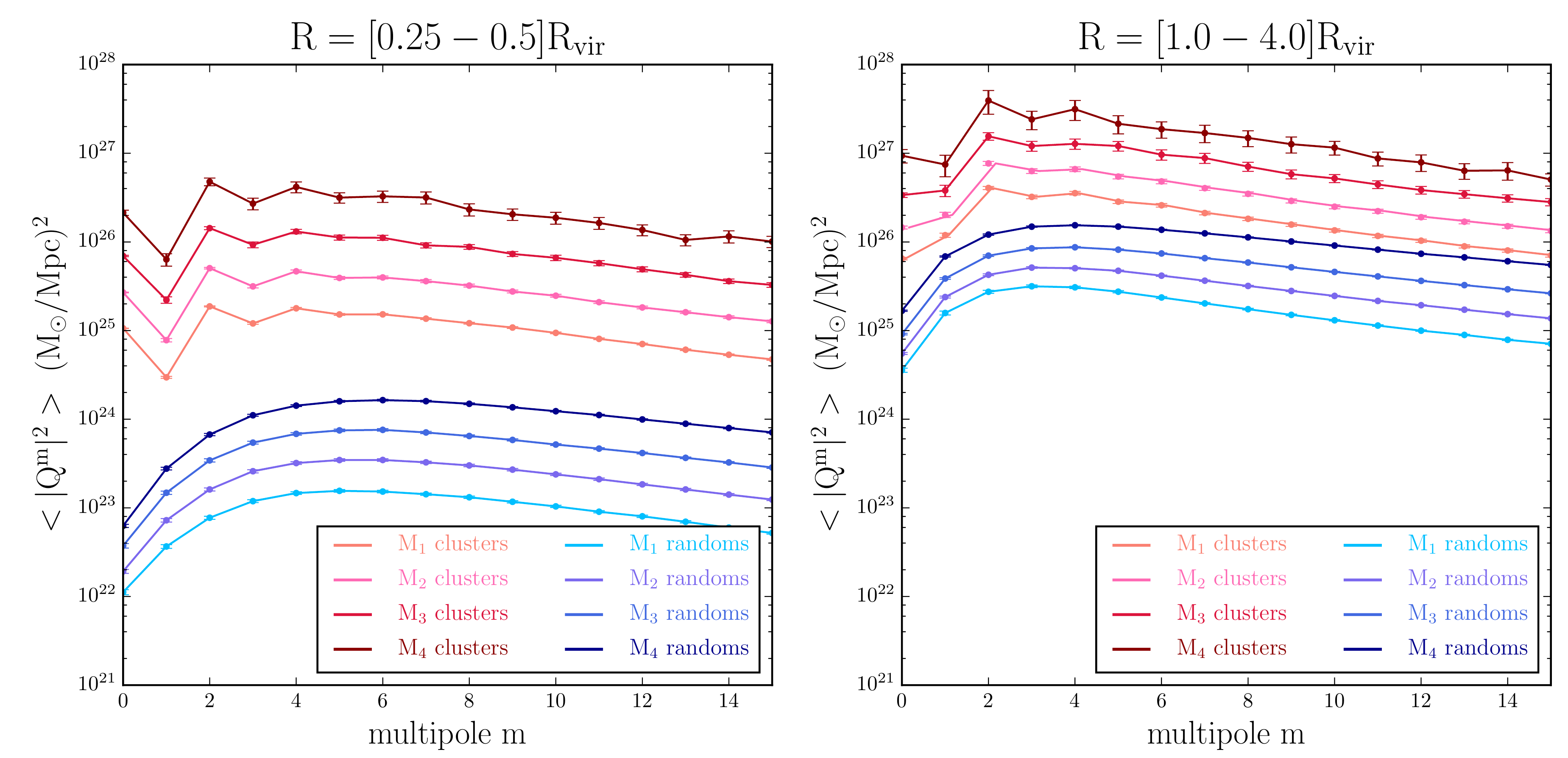}  
\caption{ Multipolar moments spectra around clusters (upper red curves) and around random location (lower blue curves)  for the two annuli $0.25 \le R/R_{\rm vir}< 0.5$  and $1 \le R/R_{\rm vir}< 4$ ({left and right-hand panels} respectively) for the $z=0$ simulation output. There is an excess of power around clusters as compared to random localisations, since the density field is denser. At large $m$ (substructures scales), their spectra differ by a nearly constant multiplicative factor $\hat{\alpha}$, which is the first order signature of the dynamical evolution of the shape of clusters (see Sect.~\ref{resultz0}).
}
\label{fig:mainres}
\end{figure*}
%%%%.........

The projected surface density of a discrete distribution of particles of mass $M_j$  reads
\begin{equation}
\Sigma (r,\varphi) = \sum_{j} M_j\; \delta_{\rm D}({\vec{r}}-{\vec{r}_j})\,,
\end{equation}
where $r$ and $\varphi$ are the coordinates in the plane of the sky.
Hence, translating Eq.~\ref{eq:mpoledef} to a  discrete distribution, the multipolar moments reads
\begin{align}
Q_m & = \int_{0}^{\infty} dr r^{m+1} w_m(r) \int_{0}^{2\pi} d\varphi \ e^{\imath  m\varphi} \ \Sigma(r,\varphi)\,, \nonumber\\
    & = \sum_j M_j  r_j^{m} w_m(r_j) e^{\imath  m \varphi_j} \,. \label{eq:mpolesum}
\end{align}
Because of the spherical extraction of particles  performed around each cluster, one needs to subtract off the contribution of the cosmological mean density, which simply reads
\begin{equation}\label{eq:sigma0}
\Sigma_0(r) = 2 \rho_{\rm mean} \sqrt{R_{\rm H}^2 - r^2}  \,.
\end{equation}
Since it does not depend on the azimuth $\varphi$, it only involves a non-zero correction $Q_0^{\rm bg}$ to the monopole term $m=0$ in Eq.~\eqref{eq:mpolesum}.
As discussed, the practical measurement of these multipolar moments requires signal stacking and, by symmetry, the phases of $Q_m$ will be lost in this process. Any departure from circular symmetry would thus be washed out. A simple workaround that does not depend on the visible baryonic mass in clusters is to consider the mean power of their multipolar moments in  harmonic space. The focus is therefore on the statistics of $\langle  \vert Q_m^2 \vert \rangle $.

For each simulated cluster $i$, the projections along the three canonical $(x,y,z)$ directions are averaged for a given annulus $\Delta R$ 
\begin{equation}
 3 \;\vert Q_m^i \vert^2 = { \vert Q_m^{i,x} \vert}^2 + { \vert Q_m^{i,y} \vert}^2 + { \vert Q_m^{i,z} \vert}^2 \,.
\end{equation}
Finally, for a given mass bin $\Delta M$, one averages the multipolar moments of haloes within the same mass bin to compute the spectrum of multipolar moments 
\begin{equation}
\langle \vert Q_m \vert^2 \rangle (\Delta M,\Delta R) = \frac{1}{N_{\rm haloes}} \sum_i^{N_{\rm haloes}\in \Delta M} \vert Q_m^i \vert^2 (\Delta R)\,.
\end{equation}
In order to explore how the background cosmology  affects  $\langle \vert Q_m \vert^2 \rangle$, these multipoles are also computed  for spheres drawn  randomly inside the  simulation box. For each position, a virial radius (and mass) is randomly assigned from the parent halo catalogue and  the multipolar moments of these random ``haloes'' is measured. 
This  allows us to contrast the growth of moments near  clusters to the overall cosmic growth of structures cast into the particular filtering of the density field as given by Eq.~\eqref{eq:mpoledef}.
In order to limit the noise in these reference moments, many more random positions than haloes are drawn. Appendix~\ref{app:checkrandom} checks that this approach yields results that are consistent with a formal integration of the power spectrum of the density contrast $P_\delta(\vec{k})$. 

At this stage, these  (centred or random) spectra are not corrected from the shot noise contribution due to the finite number of particles in the simulation that are sampling the density field. This correction is only substantial at low mass, on small scales (the smallest $\Delta R$ annulus), and for the largest multipole orders $m$. 
Shot noise corresponds to a white convergence (or surface density, here) power spectrum that is independent of the wave vector$\vec{k}$ as described in section \ref{ssec:anal2pt}
\begin{equation}
P(k) \equiv P_0 = \langle \Sigma_0(r) \rangle \, M_{\rm part}
\, ,
\end{equation}
with $\langle \Sigma_0(r) \rangle $ the mean projected density for $r \in [R_{min}, R_{max}]$.
For multipoles centred on clusters, this latter equation is multiplied by $1 + {\langle Q_0^{\rm cluster} \rangle}/{\langle Q_0^{\rm bg} \rangle }$ to get the correct shot noise amplitude. Here the average accounts for cluster-to-cluster variations of $R_{\rm max}$ (due to $M_{\rm vir}$ variations within that mass range).

\begin{figure}
\centering
\includegraphics[width=0.49\textwidth]{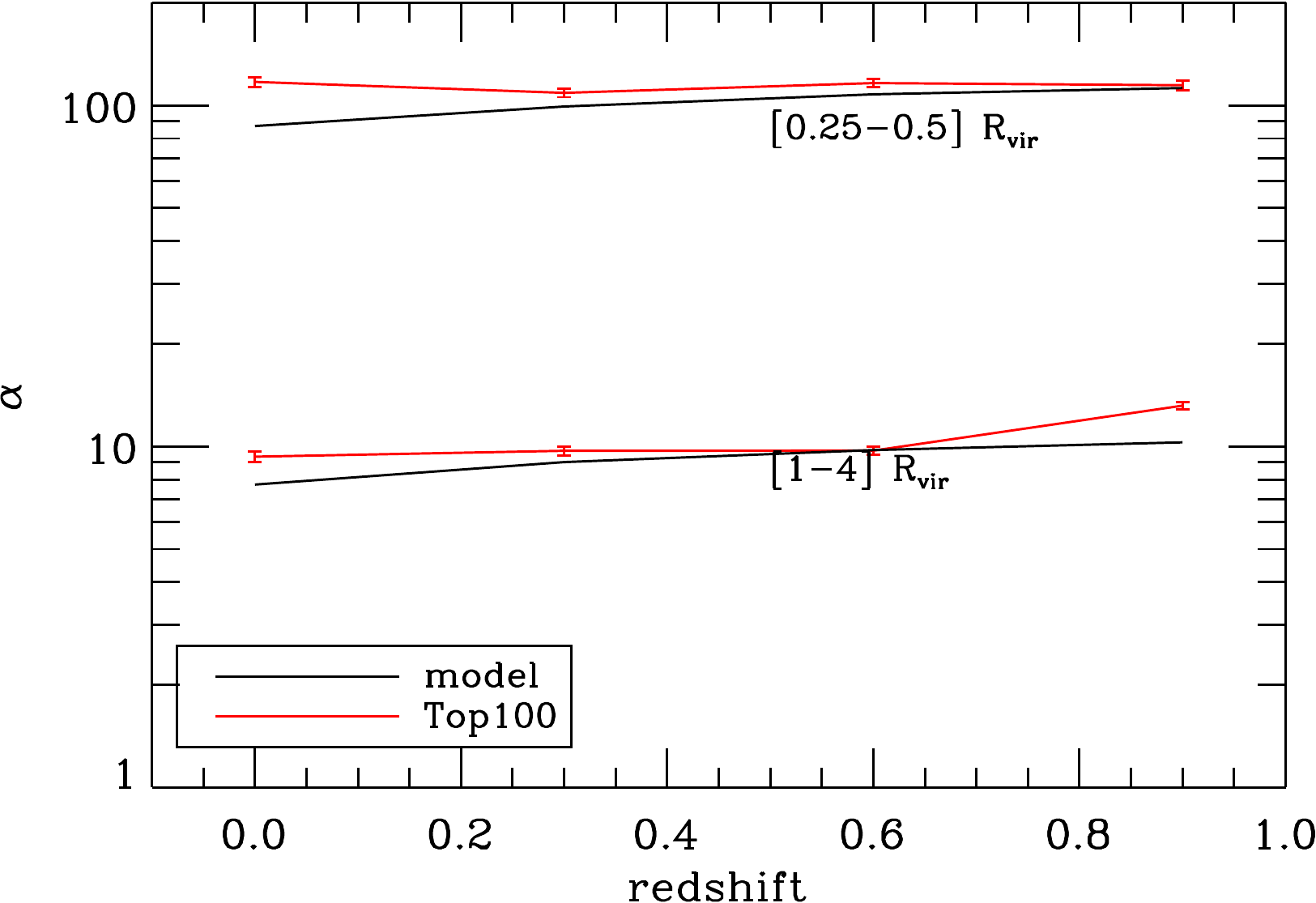}
\caption{ Evolution with redshift of the normalisation factor $\hat\alpha$ of multipolar moments spectra for the 100 most massive haloes (red solid lines). The theoretical prediction of $\alpha$ is overlaid in black. The model reproduces  the variations with time, mass and scale remarkably well, despite a crude treatment of projections and the limited validity of the spherical collapse model.
}
\label{fig:alpha_z}
\end{figure} 
Fig.~\ref{fig:mainres} presents the shape of the $\langle \vert Q_m \vert^2 \rangle$ spectrum of multipolar moments at redshift $z=0$ as a function of multipolar order $m$, and for the four mass bins and two radial bins. In order to highlight the effect that a galaxy cluster has on the statistics of $\langle \vert Q_m \vert^2 \rangle $,  we also display the same quantity  for random locations.
 
 %.......................................
\subsection{Overall excess of power}\label{resultz0}
There is an obvious excess of power in Fig.~\ref{fig:mainres} at almost all multipoles and scales around clusters as compared to random locations of similar size since the density field is denser due to the presence of the central cluster. Clusters and random locations differ by a nearly constant multiplicative factor, which is the first order signature of the dynamical evolution of the shape of clusters, as quantified in Sect.~\ref{ssec:analZel}.
 This multiplicative factor $\hat{\alpha}$ (the 2D counterpart of the boost factor $\alpha$ of Sect.~\ref{ssec:analZel}) is estimated empirically by focussing on the $m\in[15-30]$ multipole range, since it is on the smallest angular scales that the assumptions behind the Zeldovich boost are most sensible: 
\begin{equation}
\log \hat{\alpha}  =  \frac{1}{16} \sum_{m=15}^{30} \left( \log{\langle \vert Q_m \vert^2 \rangle_{\rm cluster}} -  \log{\langle \vert Q_m \vert^2 \rangle_{\rm random}}  \right)\;.
\end{equation}
Fig.~\ref{fig:alpha_z} shows the evolution of $\hat{\alpha}$ with redshift and as a function of scale for the 100 most massive clusters in the simulation box. This boost, of order $\sim100$ on the small  $R\in [0.25,0.5] R_{\rm vir}$ scale, and  of order $\sim 10$ in the larger $R\in [1,4] R_{\rm vir}$ scale, is, of course, of key importance for the detectability of $\vert Q_m\vert^2$.

A comparison with the prediction of $\alpha$ based on the spherical contraction of a Lagrangian shell  presented in Sec.~\ref{ssec:analZel} is overlaid. It shows a remarkable agreement, given the limited validity of the extension of this model deeply inside virialised haloes and the lack of modelling of projection effects. Since we use apertures that scale with the virial radius, $\alpha$ should not evolve much with time or mass (up to a mild change with time and mass of the halo concentration, see Appendix~\ref{app:zeldo}). This is clearly seen and reproduced in Fig.~\ref{fig:alpha_z}.

 %%%%.........
\begin{figure*}
\centering
\includegraphics[width=0.9\textwidth]{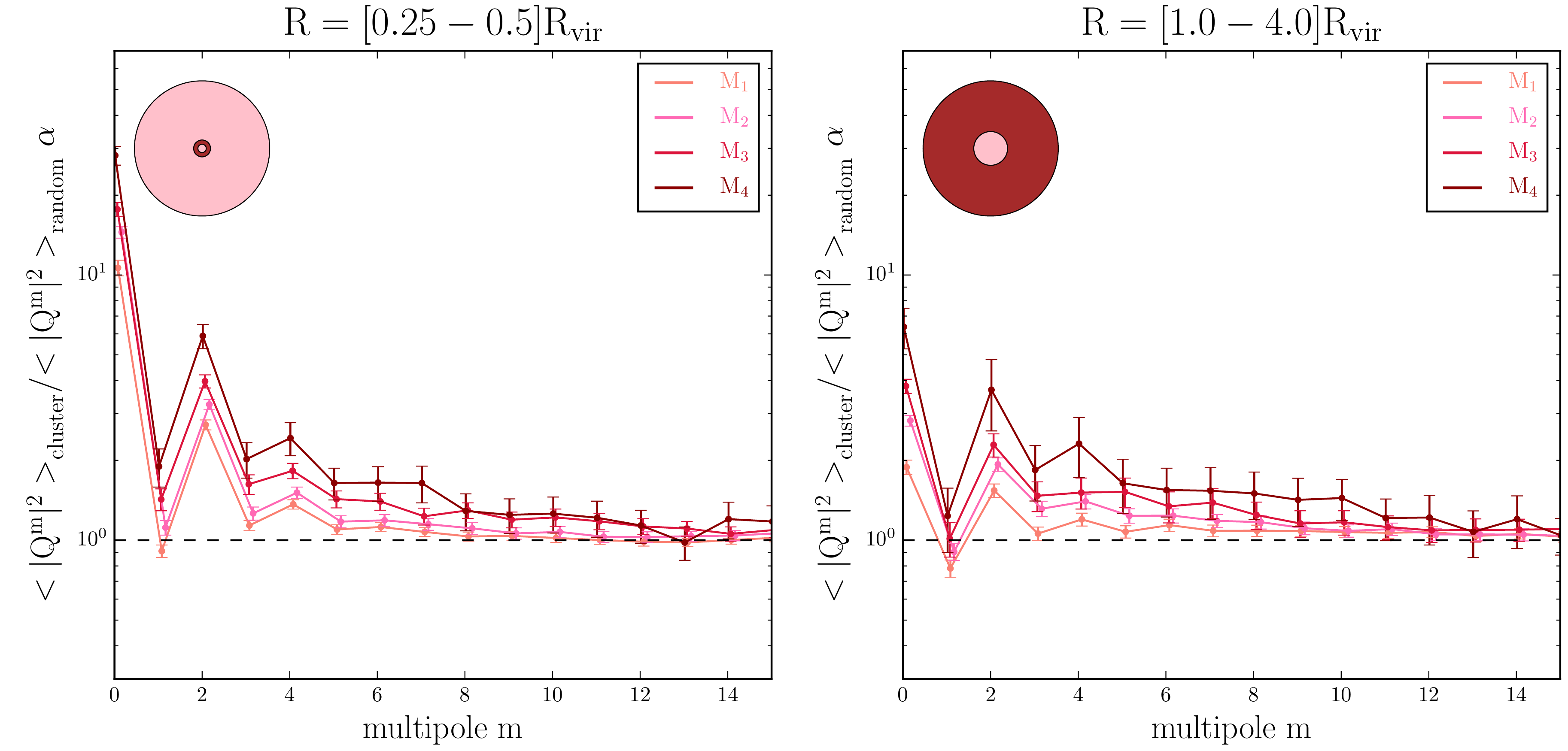}
\includegraphics[width=0.9\textwidth]{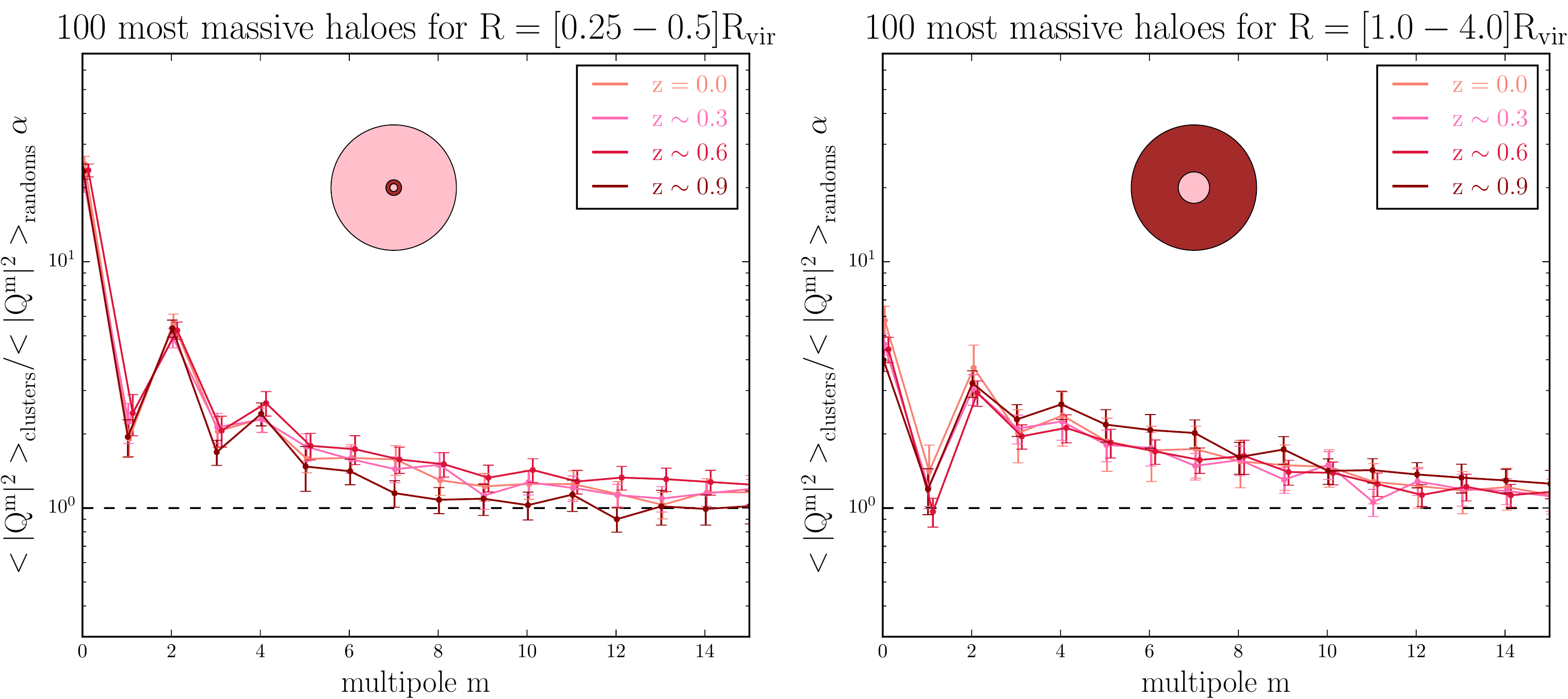}
\caption{{\it Top Panels: } Multipolar moments spectra (normalised by random profiles) for the annulus  $0.25 \le R/R_{\rm vir}< 0.5$ (left) and $1 \le R/R_{\rm vir}< 4$ (right) at redshift $z=0$ as a function of halo mass.
 In the internal region, the quadrupole prevails, and reflects the ellipsoidal symmetry of the core. In the external region, the residual power excess at $2<m<10$, not predicted by GRF approximation, should represent the nonlinear sharpening of the filamentary structure.
  {\it Bottom Panels: } Multipolar moments spectra (normalised by random profiles) for the annulus $0.25 \le R/R_{\rm vir}< 0.5$ (left) and $1 \le R/R_{\rm vir}< 4$ (right) for the 100 most massive haloes for the $z=0$, $z\sim 0.3$, $z\sim 0.6$ and $z\sim 0.9$ simulation outputs.
 Comparing a population of haloes with the same level of nonlinearity (or the same rareness), the shape of haloes appears to be settled early in the cosmic history.
 }
 \label{fig:z_evol}
\end{figure*} 

%.......................................
\subsection{Harmonic distortions}
In order to highlight spectral distortions that are in excess of the overall boost,    Fig.~\ref{fig:z_evol} shows the normalised spectra 
$\langle \vert Q_m \vert^2 \rangle_{\rm clusters}/ (\alpha \, \langle \vert Q_m \vert^2 \rangle_{\rm random}) $ as a function of halo mass at redshift zero  in the top panels, and as a function of redshift for the 100 most massive haloes in the bottom panels.

A residual excess of power  is found on large angular scales $m\lesssim 8$, possibly extending slightly further for the outermost radial bin and for more massive haloes.  Systematically, odd orders carry less power.
This should be the signature of the peak constraint (that was only affecting $m=1$ for a GRF). 
The centering  reduces the power excess at odd multipolar orders, mainly at $m=1,3$. Mis-centering will thus reduce the contrast between odd and even orders. We explored the amplitude of this effect by applying random offsets of the order $0.1\ R_{\rm vir}$, and only found noticeable differences on small-scale moment spectra whereas offsets as large as $0.5 R_{\rm vir}$ are required to substantially change moments in the $1-4 R_{\rm vir}$ range. The main effect of mis-centering on small-scale moments is to reduce the contrast between odd and even orders, leaving the latter ones almost unchanged.

Comparing small ($0.25 \le R/R_{\rm vir}< 0.5$) and large-scale ($ 1.0 \le R/R_{\rm vir}< 4.0 $) annuli, respectively, on the left and right panels of Fig.~\ref{fig:z_evol},  a  similar excess is found suggesting that the small- and large-scale shape might be correlated (see Sect.~\ref{sec:correl} below). The faster damping   with $m$ inside haloes represents a noticeable difference, tracing the higher level of symmetry in the core of virialised structures. It  can indeed  be approximated by an ellipsoid (with possibly some amount of $m=4$ boxiness). In fact, the  inner quadrupole $m=2$ presents a higher amplitude relative to the monopole than in the outskirts of clusters (right panel). This is consistent with recent studies which use elliptical or triaxial models to describe dark matter haloes shape \citep{Warren+92,Jing+02,Despali+14}. 

More massive haloes  are more sensitive  to the anisotropic environment they formed in. Interpreting this excess of power as the non-linear sharpening of the filamentary structure of haloes sitting at the nodes of the cosmic web, one  may infer that  these  haloes are connected to a larger number of filaments, as already found by \citet{Pichon+10} and \citet{Aragon+10}. 
Massive haloes are more likely to be in their early formation phase, and their shape is typically   distorted  by major mergers or accretion  along the preferred direction set by their connecting filaments. Conversely, lower-mass haloes are formed at higher redshift and  have had more time to relax. They  typically lost  the memory on their accretion history, and therefore the preferential directions induced by recent   merging  events.

Following the same haloes with time (bottom panels), that is, at constant initial overdensity  while compensating for  progenitor bias \citep[e.g.][]{shethettormen2004}, one can see that no significant evolution of the multipole is observed. This suggests that the shape of haloes is settled early in the cosmic history, probably in the  initial conditions, as anticipated in Sect.~\ref{sec:simPlus} \citep[and discussed in e.g.][]{Bond+96}.  Though one might have expected to observe a disconnection of dark haloes through the dark-energy-induced stretching  of the cosmic web \citep[][]{Pichon+10}, it turns out that the most massive clusters of the simulation have not had   time  to disconnect from the cosmic web nor fully relax. Their outskirts are still imprinted by their initial environment, whose azimuthal geometry displays power on a fairly wide range of multipoles triggered by the connected  filaments and walls.  Both large- and small-scale moments are frozen in shape from the initial conditions and only grow with time at the cosmic rate captured by the boost.
Their excess multipole seems qualitatively consistent  with the expected  number  of  connected  filaments (peaking at 2-4) inferred from the initial conditions  \citep[][]{Pichon+10}, keeping in mind that multipoles are mass weighted.

%.......................................
\subsection{Radial correlations}\label{sec:correl}
\begin{figure}
\centering
\includegraphics[width=0.45\textwidth]{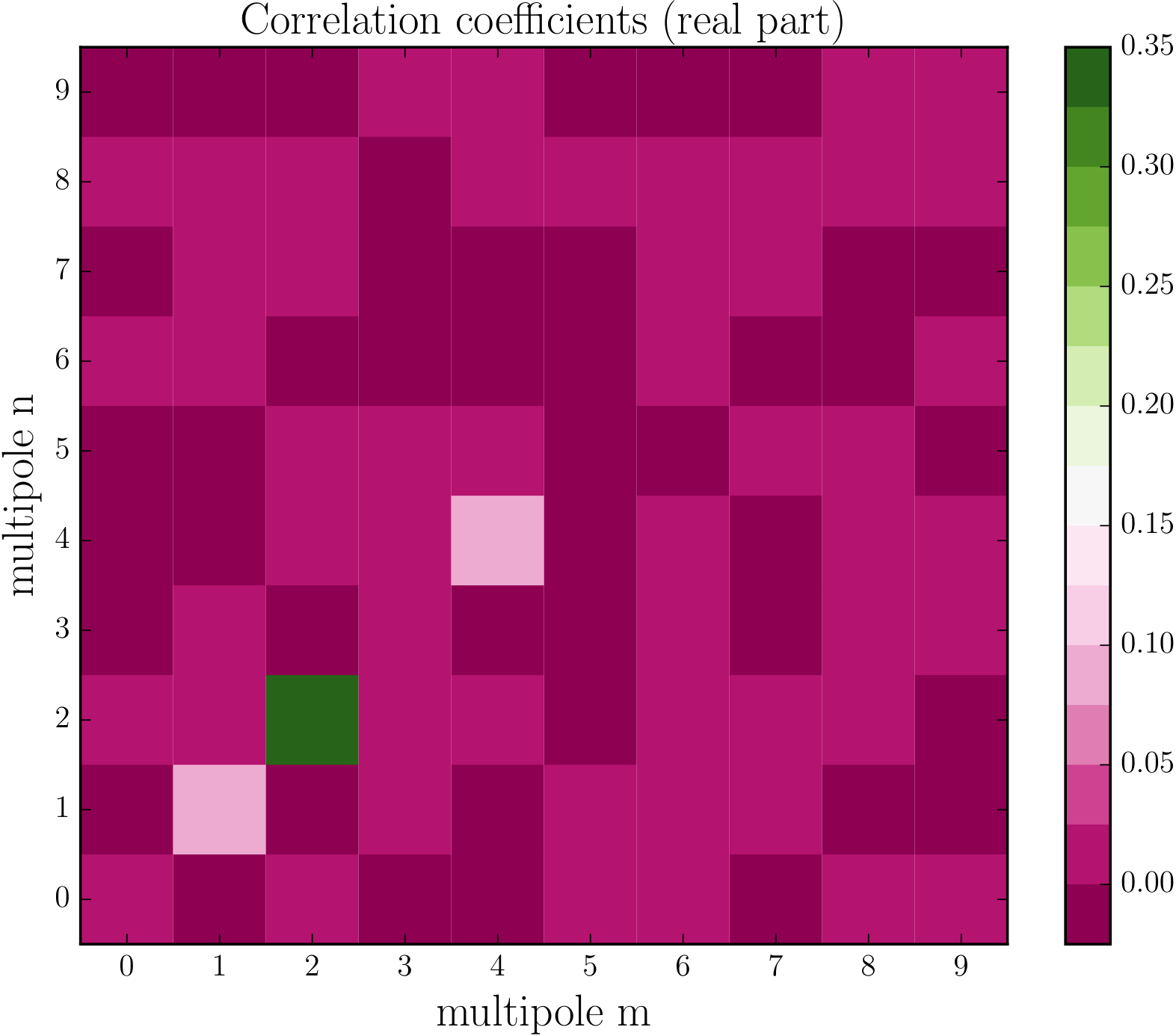}  
\caption{Cross-correlation of multipolar moments between annuli $\rm R\in [0.25,0.5] R_{\rm vir}$ and $\rm R\in [1-4] R_{\rm vir}$ for the $\rm M_1$ mass bin. Apart from $m=n=1, 2, 4$ terms, no significant correlation is found. }
\label{fig:crosscor}
\end{figure}

Let us investigate now whether the angular shape of galaxy clusters at small and large scales are correlated.
  A cross-spectrum of multipoles at varied scales would tell us  how far  the filaments  plunge into the haloes.  Fig.~\ref{fig:crosscor} shows the reduced cross-spectrum of multipolar moments at radii $R\in [0.25-0.5] R_{\rm vir}$ and $R\in [1-4] R_{\rm vir}$ for the $M_1$ mass bin. This reads
\begin{equation}
\rho_{1,4}(m,n) = \frac{\langle (Q_m(R_1) Q_n^*(R_4) \rangle - \delta_{m0}\delta_{n0} \langle Q_0(R_1)\rangle\langle Q_0(R_4)\rangle }{\sigma_{Q_m(R_1)} \ \sigma_{Q_n(R_4)}}  \,.
\end{equation}
 The disconnected part (product of the means) is subtracted to  highlight the relative fluctuations between annuli and multipolar orders. In contrast to what was done for the auto-spectra $\langle \vert Q_m \vert^2 \rangle$ for which the mean $\langle Q_0 \rangle^2$ was not subtracted off because it contributes to the overall signal amplitude and to its detectability (Sect.~\ref{sec:snr}). Without subtraction, $\langle Q_0(R_1)\rangle\langle Q_0(R_4)\rangle$ would induce a large correlation $\rho_{1,4}(0,0) \ge 0.5$.
 
 Except for the diagonal $m=n=1, 2, 4$ terms, no significant correlation is observed. A similar trend is found at higher masses although the signal to noise is even lower. Apart from the quadrupole, the shapes seem to decorrelate. 
Hence there is no strong angular coherence between the structures found at $R\in [0.25-0.5] R_{\rm vir}$ and $R\in [1-4] R_{\rm vir}$ beyond the quadrupole.
The non-zero cross-correlation at $m=1$ is induced by the condition to be centred on a density peak. We also checked that miscentring has the effect of reducing the $\langle Q_1(R_1) Q_1^{*}(R_4)\rangle$ term. In order to reach a factor 2 decline in correlation amplitude, one needs to reach offsets of the order $0.2\,R_{\rm vir}$.

%%%%.........

%------------------------------------------------------------------------------ 
%------------------------------------------------------------------------------ 
\section{Measuring multipolar moments from shear data}\label{sec:snr}

Let us now quantify our ability to estimate the power spectrum of multipolar moments from real WL data. They involve a set of background galaxies whose light is deflected by a foreground galaxy cluster.  Since very few detections of filaments, and little evidence for a departure from circular symmetry have been reported so far,
there is a dire need to investigate how  to measure the mean excess power spectrum of multipolar moments from a set of galaxy clusters. The aim is,   in particular, to quantify  the  expected    signal to noise  ratio as a function of redshift, scale, and mass.

Three leading sources of noise must be considered here: The shape noise coming from the unknown intrinsic ellipticity of background sources, the lensing gravitational potential due to the LSS along the line-of-sight, and the sample variance due to the finite number of clusters one can average over to infer the mean multipolar spectrum. One expects some internal variability due to the relative dynamical age of these massive structures undergoing merger, being  fed by a variable number of filaments. The derivation we use follows \citet{schneider98}  who studied the statistical properties of the aperture mass $M_{\rm ap}$ estimators, but neglects the effects of Poisson fluctuations in the number of background sources. This is legitimate for large-scale cluster lensing with deep wide-field imaging.

One first needs to rescale the multipolar moments previously inferred from simulations in terms of mass per Mpc since the convergence was replaced by the surface density in equation~\eqref{eq:mpolesum}. Therefore the  moments will be multiplied by the mean inverse critical density $\langle \Sigma_{\rm crit}^{-1}(z) \rangle $, averaged over the same redshift distribution of background sources as that assumed in Sect.~\ref{ssec:anal2pt}. 
 The corresponding values for the fiducial lens redshifts are listed in Table~\ref{tab:scritvals}.

%.............
\begin{table}{\small
\center\begin{tabular}{lcccc}\hline\hline
\T
  lens redshift  & 0 & 0.3 & 0.6 & 0.9 \\
  $\langle \Sigma_{\rm crit}^{-1}\rangle^{-1} \;\;  [10^{15}\,\msun\,{\rm Mpc}^{-2}]$ & $\infty$ &  3.330 & 4.413 & 7.585 \B
 \\
   \hline\hline
\end{tabular}
\vskip 0.25cm
\caption{Critical lensing surface density used to convert Sect.~\ref{sec:simus} multipolar moments expressed in terms of mass into moments expressed in terms of convergence.}
\label{tab:scritvals}}
\end{table}

%.......................................
\subsection{Covariance of multipolar moment spectrum estimators}

The convergence field around a cluster should be replaced by three uncorrelated fields
\begin{equation}
  \kappa \rightarrow  \kappa + \kappa^{\rm s} + \kappa^{\rm L} \,,
\end{equation}
where $\kappa$ is the signal produced by the cluster, $\kappa^{\rm s}$ is the contribution from the intrinsic ellipticity of background sources and $\kappa^{\rm L}$ is due to the uncorrelated LSS along the line-of-sight.
The LSS that is not physically correlated to the direct cluster environment (beyond 10 Mpc, or so) will nevertheless give rise to a cosmological convergence field that will act as an additive source of noise plaguing the multipolar moment (or more generally the convergence field sourced by the clusters)  to be measured. This so-called cosmic shear signal has to be taken into account for the detectability of multipolar moments  as it is a substantial source of noise for overall cluster-mass measurements \citep{hoekstra01b,Hoekstra+03}. Sect.~\ref{ssec:anal2pt} already showed the two-point properties of this component $\kappa^{\rm L}$ through the equation \eqref{eq:mpolecov}. In the remainder,   the statistical properties of $\kappa^{\rm L}$ are approximated as those of a GRF (with null kurtosis) since the focus is on the statistics of $\kappa$.

The convergence field is not directly observable; ellipticities are used to measure shear.
The observed complex ellipticity $\epsilon$ of a galaxy with intrinsic source plane ellipticity $\epsilon_{\rm s}$ and carrying a complex shear signal $\gamma$ is simply: $\epsilon = \epsilon_{\rm s} + \gamma$. Since for an ensemble of galaxies with random intrinsic orientation, the mean $\epsilon_{\rm S}$ is null, one can write:
\begin{eqnarray}
\langle \epsilon_i \rangle & = & \gamma_i\,, \\
\langle \epsilon_i \epsilon_j \rangle & = & \gamma_i \gamma_j + \gamma^{\rm L}_i \gamma^{\rm L}_j + \sigma_\epsilon^2 \delta_{ij}\,,
\end{eqnarray}
where  the one-dimensional rms dispersion of intrinsic source ellipticities $\sigma_\epsilon \simeq 0.25$ was introduced. In practice, measurement errors that depend on the quality of images (correction for smearing by the Point Spread Function, signal-to-noise...) would also increase this dispersion to a value that we shall take to be $\sigma_\epsilon \equiv 0.3$ in the remainder. 

Following again \citet{SB97}, let us consider the local estimator of $Q_m$ from measured shear as defined by Eq. \eqref{eq:mpoledef2l} and in which only background galaxies projected into a given annulus of inner and outer radius $\nu \rmax$ and $\rmax$ are used to estimate the multipolar moments in that aperture. The non-local estimator, involving shear measurements outside that aperture, may increase the sensitivity but at the expense of introducing large correlations between annuli.  The remainder of this work only explores the merits of the local estimator \eqref{eq:mpoledef2l} that  is rewritten in the same way as \citet{SB97}
\begin{equation}\label{eq:mpoleREdef}
  Q_m  = \int\der^2\vec{r} \left[\frac{ b_{t,m}(r)}{r}  \gamma_t(r,\varphi)   +  \imath  \frac{b_{\times,m}(r)}{r} \gamma_\times(r,\varphi) \right] \ {\rm e}^{\imath  m \varphi}\,,
\end{equation}
with
\begin{eqnarray}\label{eq:bdef}
b_{t,m}(r) & = & r^{m+1} w_m(r) \,, \\
b_{\times,m}(r) &= & r^{m+1}  \left(w_m(r) + \frac{r}{m} w'_m(r) \right) \,.
\end{eqnarray}
For a finite number $N$ of sources inside a given annulus $[\nu\rmax,\rmax]$, the discrete version of \eqref{eq:mpoleREdef} reads
\begin{equation}\label{eq:mpoleREdefDISC}
 \hat{Q}_m = \frac{1}{\overline{n}}  \sum_k^N {\rm e}^{i m \varphi_k}  \left[\frac{ b_{t,m}(r_k)}{r_k}  \epsilon_{{\rm t},k}   
              +  \imath \, \frac{b_{\times,m}(r_k)}{r_k} \epsilon_{\times,k} \right] \,,
\end{equation}
where $\overline{n}$ is the mean number density of background sources, for which a typical value $\overline{n} = 30\, {\rm arcmin}^{-2}$.
Even though the sources are randomly oriented, the ensemble average of the quadratic estimator $ \vert \hat{Q}_m \vert ^2 $ of
$ \vert Q_m \vert ^2  $ will contain a shape noise and a LSS noise contribution that must be subtracted off
\begin{equation}\label{eq:mpoleM2}
\langle \vert \hat{Q}_m \vert^2  \rangle = \vert Q_m \vert^2  + \vert Q_m^{\rm s} \vert^2 + \vert Q_m^{\rm L} \vert^2\,.
\end{equation}
As shown in \citet{SB97}, this shape noise mean power spectrum is 
\begin{eqnarray}\label{eq:shapenoisemean}
\vert Q_m^{\rm s} \vert^2  &= & \frac{\sigma^2_{\epsilon} }{\overline{n}^2 }  \sum_k^N \frac{ b_{t,m}^2(r_k) + b_{\times,m}^2(r_k) }{ r_k^2}\,,\\
 &= & \frac{\pi \sigma^2_{\epsilon}}{\overline{n} } \int_{\nu R}^R dr \frac{b_{t,m}^2(r)+b_{\times,m}^2(r)}{r}\,.
\end{eqnarray}

This mean  noise spectrum gives us a sense of our ability to measure multipolar moments. But one really needs to compute the covariance of the estimator by following the calculations  made for cosmic shear correlation functions \citep{schneider98,Schneider02} and consider the covariance matrix of the multipolar moments
\begin{equation}\label{eq:covariance}
\begin{split}
\sigma^2_{mn} & \equiv &\langle \vert \hat{Q}_m \vert^2 \vert \hat{Q}_n \vert^2\rangle - \langle \vert \hat{Q}_m \vert^2  \rangle \langle \vert \hat{Q}_n \vert^2  \rangle, \\
             & = &    \delta_{mn} \vert Q_m^{\rm s} \vert ^2  \left[ \vert Q_m^{\rm s} \vert ^2  + 2 \vert Q_m \vert^2  + 2 \vert Q_m^L \vert^2 \right] +  \\
           &&  \delta_{m0}\delta_{n0}  \vert \tilde{Q}_0^{\rm s}\vert^2   \left[ \vert\tilde{Q}_0^{\rm s}\vert^2    + 2  \vert Q_0\vert^2  + 2 \vert Q_0^L\vert^2  \right] +\\
          && 4 A_{\kappa,mn}  A_{\kappa^{\rm L},mn}  + \\
          && \langle \vert Q_m \vert^2 \vert Q_n \vert^2\rangle - \langle \vert Q_m \vert^2  \rangle \langle \vert Q_n \vert^2 \rangle + \\
          && \langle \vert Q^{\rm L}_m \vert^2 \vert Q^{\rm L}_n \vert^2\rangle - \langle \vert Q^{\rm L}_m \vert^2  \rangle \langle \vert Q^{\rm L}_n \vert^2 \rangle \phantom{+} \;.
\end{split}
\end{equation}
The first two terms in \eqref{eq:covariance}, containing $Q_m^{\rm s}$, correspond to the shape noise. They are diagonal and dominate for $m=0$. Their derivation along with the definition of the modified moments $\vert \tilde{Q}_m^{\rm s} \vert^2$ are detailed in Appendix \ref{app:covar}. 
The last three terms correspond to the mixture of sampling variance and LSS noise contributions. If both $\kappa$ and $\kappa^{\rm L}$ were GRFs, these three terms would simplify to $2 ( A_{\kappa,mn} + A_{\kappa^{\rm L},mn})^2$, with $A_{mn}$ defined in Eq.~\eqref{eq:mpolecov}.

%.......................................
\subsection{Overall detectability}

%..................
\begin{figure*}
\centering
\includegraphics[width=0.95\textwidth]{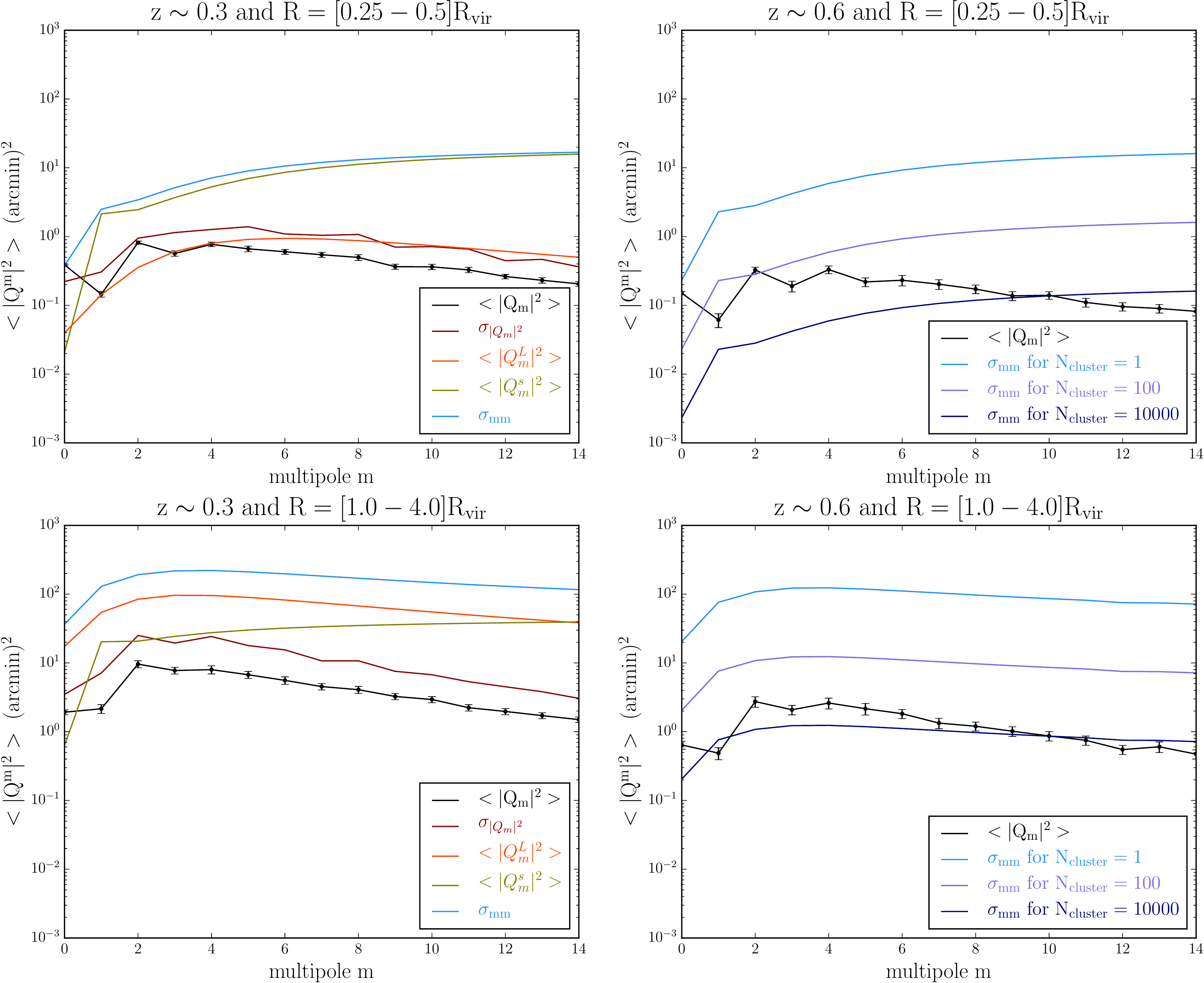}
\caption{Multipolar moment spectra of clusters for the mass bin $\rm M_3$ and the noise contribution at two scales in the annuli $ \rm \Delta R_1$ (top panel) and  $\rm \Delta R_4$  (bottom panel) and for two redshift values $\rm z=0.3$ and 0.6, respectively, from left to right. The left-hand panels show the different sources of noise as described in Sect.~\ref{sec:snr}. On the right-hand panels,  the total noise profile $\sigma_{mm}$  is shown  once divided by $\sqrt{N_{\rm clusters}}$ in order to mimic the shot noise obtained by  stacking the spectra of $N_{\rm clusters}$ clusters.}
\label{fig:qnnoise}
\end{figure*}

 The left-hand panels of Fig.~\ref{fig:qnnoise} compare the amplitude of $\vert Q_m\vert^2$, $\vert Q_m^{\rm s}\vert^2$ and $\vert Q_m^{\rm L}\vert^2$ along with the rms dispersion of $\vert Q_m \vert^2$ for a single cluster of galaxies in the intermediate mass bin ($ M \in [4-8] \times 10^{14} \hmmsun$)  at redshift $ z \sim 0.3$. The total combination of diagonal terms $\sigma_{mm}$ in Eq.~\eqref{eq:covariance} is also overlaid. It appears that for the internal regions (top panel), the shape noise -- coming from the intrinsic ellipticities of the background galaxies --  dominates, whereas on large scales, the dominant source of noise is the line-of-sight density fluctuations $\vert Q_m^{\rm L}\vert^2$. 
\citet{Hoekstra+03} also found a similar radial behaviour of the relative importance of shape noise and LSS for measuring overall cluster masses.
Obviously,  detectability is easier in the central region than at the outskirts  of clusters, since the density field is stronger. The signal-to-noise declines also as redshift increases and mass decreases.
The overall signal-to-noise is quite low for a single cluster, with values of order $0.1$ for the most favourable multiples ($m=0,2,4$). On small scales, using a non-local multipole estimator might slightly reduce shape noise contributions, but on large scales, LSS will dominate anyway, suggesting that very deep observations with a low shape noise level would not be of much help. We must therefore consider stacking the lensing signal of many clusters. 

On the right panels of Fig.~\ref{fig:qnnoise}, the amplitude of $\langle \vert Q_m\vert^2 \rangle $ is compared to the total noise contribution $\sigma_{mm}$ for a single cluster in the mass bin $ \rm M_3$ at $z \sim 0.6$ at small and large radii. Instead of splitting the noise budget into its components, the total overall noise level is scaled by a factor 0.1 or 0.01 as it should naturally decrease if one considers 100 or $10\,000$ clusters instead of just one. Besides,  at higher redshift, the signal-to-noise decreases due to the rise of the critical density with redshift (see Table~\ref{tab:scritvals}).

\begin{figure*}
\centering
\includegraphics[width=0.95\textwidth]{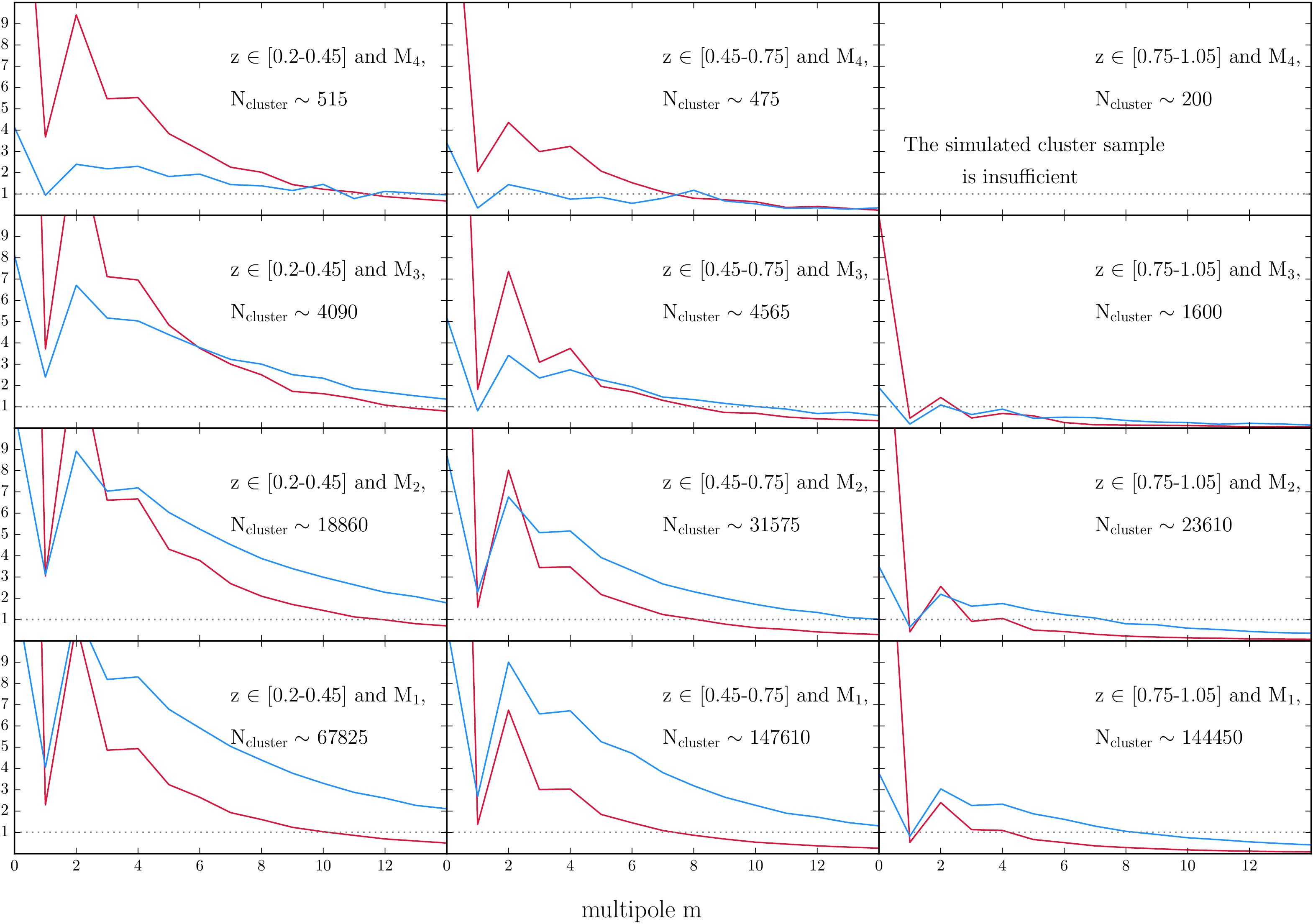}
\caption{Expected signal-to-noise ratio of multipolar moment spectra that one could obtain by stacking the total number of galaxy clusters $N_{\rm cluster}(z,M)$, that will be uncovered in the future \textit{Euclid} survey \citep{Sartoris++16}. Red and blue curves represent the SNR for the annuli $R_1 \in [0.25-0.5] R_{\rm vir}$ and $R_4 \in [1-4] R_{\rm vir}$, respectively.  Three different ranges of redshift are considered ($\sim 0.3,\ \sim 0.6$ and $\sim 0.9$ from left to right), together with four fiducial cluster mass intervals ($M_4,\cdots M_1$, from top to bottom).}
\label{fig:snr_euclid}
\end{figure*}

The \textit{Euclid} photometric galaxy cluster survey will contain about $2\times 10^{5}$ clusters between $z=0.2$ and $z=2$ \citep{Sartoris++16}. The authors provide an estimate of the number of galaxy clusters to be detected for a given range of redshift and minimum mass by carefully accounting for the {\it Euclid} cluster selection function. This allows us to predict the expected number of clusters   in each   mass bin and for three redshift intervals, hence the expected signal-to-noise ratio on multipolar moments. This is shown in Fig. \ref{fig:snr_euclid} for both internal and external regions of galaxy clusters.
Higher signal-to-noise can be achieved for lower mass clusters because of their larger abundance.
  More specifically, multipolar moments measured at the outskirts of clusters should accurately be detected by stacking clusters with $M_{\rm vir}\le 8 \times 10^{14} \hmmsun $ and $z \le 0.75 $. In the internal regions, we estimate that the angular symmetries on cluster cores could be probed for all cluster masses for $z \lesssim 0.75 $.

In order to improve the signal-to-noise and permit detections on a shorter timescale, one can consider a broader annulus $\rm R =[0.1-1.0] \ R_{\rm vir} $ that probes the high-density regions, the corresponding list of signal-to-noise ratios for 100 clusters in Table~\ref{tab:snr_R11}. A detection of the multipoles from $m=0$ to $m=4$ is possible in this annulus, stacking the signal over about 100 galaxy clusters, for all mass bins at $z \sim 0.3$ and for the most massive clusters. At $z\sim 0.6$, the measurement is possible for clusters of similar mass but only up to the quadrupole.

%..........
\begin{table}
\centering
{\small
\begin{tabular}{cccccccc}\hline\hline
\small \T
  Redshift & Mass bin & \multicolumn{6}{c}{multipole m} \\ \small
   &  & 0 & 1 & 2 & 3 & 4 & 5 \\ \hline
     \multirow{4}{*}{0.3} 
     \T
   & $ \rm M_1 $  &   9 & 0.6 & 1.8 & 0.4 & 0.4 & 0.2  \\
   & $ \rm M_2 $  &  12 & 1.2 & 3.4 & 1.0 & 1.0 & 0.5 \\
   & $ \rm M_3 $  &  14 & 2.4 & 5.4 & 2.1 & 2.2 & 1.1 \\ 
   & $ \rm M_4 $  &  11 & 3.2 & 7.1 & 3.6 & 3.7 & 2.1 \\ \hline
   \T
   \multirow{4}{*}{0.6} 
   & $ \rm M_1 $  & 8  & 0.3 & 0.9 & 0.2 & 0.2 & 0.08 \\
   & $ \rm M_2 $  & 10 & 0.6 & 2.0 & 0.5 & 0.5 & 0.2 \\
   & $ \rm M_3 $  & 12 & 1.4 & 3.7 & 1.0 & 1.1 & 0.4 \\ 
   & $ \rm M_4 $  & 17 & 3.1 & 4.9 & 2.0 & 2.4 & 0.8 \\ \hline
   \T
   \multirow{3}{*}{0.9} 
   & $ \rm M_1 $  & 5  & 0.1 & 0.3 & 0.01 & 0.01 & 0.003\\
   & $ \rm M_2 $  & 8  & 0.2 & 0.9 & 0.2 & 0.2 & 0.06 \\
   & $ \rm M_3 $  & 10 & 1.0 & 1.9 & 0.5 & 0.5 & 0.24\\ 
   \hline\hline
\end{tabular}
\vskip 0.25cm
\caption{Signal-to-noise ratio on multipolar moment spectra for $\rm N_{\rm cluster}=100$ when the annulus $R \in [0.1,1.0]\,R_{\rm vir}$ is considered. }
\label{tab:snr_R11}}
\end{table} 
%..........

%------------------------------------------------------------------------------ 
%------------------------------------------------------------------------------ 
\section{Summary \& conclusions}\label{sec:conclusion}
The  multipolar moments of the convergence were used to quantify asymmetries in the projected density field around galaxy clusters. The multipoles were computed within annuli centred on mock clusters of galaxies, extracted from a large dark matter simulation. The power spectra of these moments, $ \langle \vert Q_m \vert^2 \rangle$, were studied in detail, noticeably via their evolution with redshift, cluster mass and radial aperture. 

We quantified  the degree of angular symmetries around clusters that is in excess of the background density field. 
To this end, we compared the multipolar moment spectra centred on clusters to those of random locations.
To first order, the non-linear evolution of mass shells sinking towards the centre of clusters induces a boost of power at all angular scales due to the contraction of the Lagrangian patch initially encompassing the mass fluctuations.
When the density field is nearly Gaussian, only  $m\le2$ moments are affected by the peak constraint, as demonstrated in  \citet{Codis++17},
while in the quasi-linear regime this paper also predicts, perturbatively, the achromatic boost. In the highly nonlinear regime,
 a simple model based on the spherical collapse model was presented, and is found to be in good agreement with  measurements on massive haloes extracted from the PLUS Dark Matter cosmological simulation. 

Looking at the harmonic distortions occurring on top of the overall boost, we found  an angular power excess at orders $m\lesssim10$ in the outer shell, which traces the  azimuthal shape of the projected density.  The excess chromatic power in the multipoles seems qualitatively consistent with the predicted number of connected filaments  in   the outskirts of clusters \citep[][Pogosyan et al. in prep]{Pichon+10},
 keeping in mind that the harmonic analysis is mass weighted.
We also found a higher amplitude for the quadrupole in the central regions, 
which reflects the ellipsoidal symmetry of the core of haloes \citep[see also, e.g.][]{Despali+14}.
Given the similarity  in the  excess power in  internal and external regions, we examined the cross-correlation of multipolar moments between these two annuli. Except for the quadrupole, angular shapes at small and large scales seem to be uncorrelated.
This is probably because, typically, two branches of filaments are connected to a node of the cosmic web on small scales \citep{pogo09}.
Further away from the nodes, bifurcation points appear and therefore increase the number of filaments.
We therefore expect the quadrupole to be correlated between small and large scales but correlations beyond the quadrupole to be suppressed on small scales.
Therefore, the steeper profile of the small-scale multipoles compared to their large-scale counterpart is fully consistent with this idea that the cosmic connectivity is smaller on small scales.

We also studied the evolution of multipolar spectra at different redshifts (from $z=0$ to $z\sim0.9$) and found that following the same population of haloes, that is, with the same initial rareness level, the multipolar moments measured in an annulus that follows the growth of the virial radius grow at a rate that is similar to the cosmic rate (implying no change of the spectra normalised by $\alpha$).  This suggests that the larger-scale shape in the vicinity of the halo is fixed at early times in its formation history, and keeps the memory of its initial conditions during collapse \citep{Bond+96}. Indeed,  the  dark-energy-induced disconnection  from the cosmic web  \citep{Pichon+10} has not yet occurred for this most massive population of clusters. 
We note that, as expected, for a less massive population of haloes, we observe a decrease of the harmonic excess in both internal and external regions with time. This decrease is due to a virialisation of the core and a disconnection of the halo from the cosmic web, at the respective scales.
Globally, these quantitative estimates are consistent with dynamical expectation drawn from the visual inspection of simulations.

Finally, we estimated the detectability of these harmonics using WL data, taking into account different sources of noise such as the shape noise (intrinsic ellipticity of the background galaxies), the impact of the LSS along the line-of-sight, and  the sample variance. As expected, the amplitude of the signal-to-noise ratio increases with halo mass and depends on the aperture and the cluster redshift (see Fig.\ref{fig:qnnoise}).  On small scales (within the virial radius), shape noise dominates whereas additional deflections due to matter along the line-of-sight  dominate the noise budget on larger scales (outside the virial radius).
Due to the weakness of the signal, one has to stack the multipolar moment spectra over a larger number of clusters. Hence, one should  consider current detections of filaments with WL with caution. 

With the upcoming Euclid mission \citep{Euclid+11}, multipolar moment spectra will be detected with a good degree of precision up to $ m \sim 10$ in central and external regions.
On a shorter timescale, considering a broader annulus ($R=[0.1-1]R_{vir}$), harmonic components should be measured at orders $m=0,2,4$, by stacking $\sim 100$ massive clusters up to $z \sim 0.6$.
Such measurements on ground-based observations will  be presented in a forthcoming companion paper (Gavazzi et al, in prep.).  
Upcoming investigations should extend this study to varying cosmological models. Cosmology can have an  impact on multipolar moment spectra but the overall amplitude is governed by the cluster mass profile, which is better constrained by a direct stacking of the tangential shear or a stacking of $Q_0$ values. Cosmological effects related, for instance, to the growth rate and dark energy, may have an effect on the relative amplitude of moments. 
One could extend those measurements to cosmological hydrodynamical simulations in which  baryonic physics, such as gas cooling and feedback from active galaxy nuclei, will substantially change the shape of the total mass distribution on small scales \citep{Teyssier+11,Suto+17}.
These scales may be more efficiently  probed by strong lensing observables \citep[following, e.g.][]{peirani08}. 
Additionally, models involving warm or self-interacting dark matter may leave an interesting footprint on the spectrum of multipolar moments, as departure from vanilla CDM would tend to make haloes rounder. We thus expect these spectra to be a valuable Dark Matter probe on small scales.
Finally, we expect that comparisons between multipolar moments measured with lensing and similar moments measured on different populations of galaxies of different types may shed new light on the relative bias of these populations inside filaments and as they sink into cluster haloes. The key role of environment on the quenching of star formation in galaxies around clusters may hence be probed.

%------------------------------------------------------------------------------ 
%------------------------------------------------------------------------------ 
\begin{acknowledgements}
The authors would like to thank K. Benabed, J-F.~Cardoso, M.~Kilbinger, G. Lavaux for valuable discussions.
This work was supported by the Agence Nationale de la Recherche (ANR)
as part of the SPIN(E) ANR-13-BS05-0005 and Jeune Chercheur AMALGAM projects, and by the Centre National des Etudes Spatiales (CNES).
This work has made use of the Horizon Cluster hosted by the Institut d'Astrophysique de Paris. We thank S.~Rouberol for running 
the cluster smoothly  for us.
\end{acknowledgements}

%------------------------------------------------------------------------------ 
%------------------------------------------------------------------------------ 
\bibliographystyle{aa}
%\bibliography{references}

%------------------------------------------------------------------------------ 
%------------------------------------------------------------------------------ 
\appendix
%\addcontentsline{toc}{chapter}{APPENDICES}
%------------------------------------------------------------------------------ 
%------------------------------------------------------------------------------ 

%------------------------------------------------------------------------------ 
%------------------------------------------------------------------------------ 
\section{Towards the nonlinear statistics of $Q_m$}\label{app:zeldo}
Let us first consider the  weakly nonlinear regime for the statistics of $Q_m$  around clusters before turning to its strong nonlinear counterpart.

\subsection{The weakly nonlinear regime}
For simplicity, let us first focus on the statistics of density fluctuations in a 3D shell of matter located at a radius $r$ from the centre of a cluster of galaxies. This shell is falling onto the cluster at a rate that can be derived from the spherical collapse model or the Zeldovich approximation. It was originally located at a Lagrangian radius $q$. Specifying the initial potential $\psi(\vec{q})=\psi^{\rm l}(\vec{q})+\psi^{\rm c}(\vec{q})$ , where $\psi^{\rm c}$ is the potential generated by the peak at the centre and $\psi^{\rm l}$ is due to matter fluctuations in the original shell, the mapping between the Eulerian coordinate $\vec{r}$ and its corresponding Lagrangian position $\vec{q}$ is 
\begin{equation}\label{eq:zeldo0}
\vec{r} = \vec{q}  - D' \vec{\nabla}_q\psi(\vec{q})\,,
\end{equation}
with $D'= {D(a)}/{4\pi G \bar{\rho} a^3}$, $D(a)$ the linear growth rate, and $\bar\rho$ the mean comoving cosmic density.
The evolved density contrast is given by the Jacobian of this transformation
\begin{equation}\label{eq:zeldo1}
1 + \delta = \left\vert \delta_{ij}  - D' \psi_{,ij}^{\rm c} - D' \psi^{\rm l}_{,ij} \right\vert^{-1}\,,
\end{equation}
with $\left\vert \,\,\,\right\vert$ the determinant of its argument. Taylor expanding this relation around small values of $\psi^{\rm l}$ and defining the distortion tensor $\Gamma_{ij} = \delta_{ij} - D' \psi^{\rm c}_{,ij}$, allows us to rewrite Eq.~\eqref{eq:zeldo1} as
\begin{equation}\label{eq:zeldo2}
1 + \delta = \left\vert  \Gamma_{ij}  \right\vert^{-1} \, \left( 1 +  D'  \mathrm{Tr}(  \Gamma_{ij}^{-1}  \psi_{,ij}^{\rm l}  ) \right), 
\end{equation}
where $\mathrm{Tr}$ is the trace of its argument.
Accounting now for  the spherical symmetry of $\psi^{\rm c}$, the $\Gamma$ matrix reads
\begin{equation}
\Gamma  =  \mathrm{Diag}\left(  1- D' \psi^{\rm c}_{,qq} ,  1- D'  \frac{1}{q}\psi^{\rm c}_{,q} , 1- D'\frac{1}{q} \psi^{\rm c}_{,q} \right)  ,
\end{equation}
in spherical coordinates. Let us now also neglect the anisotropy of $\Gamma$ by assuming that the radial compression of fluctuations occurring as the shell shrinks does not significantly depart from the angular compression, so that $\Gamma_{ij}= (1- D'  \frac{1}{q}\psi^{\rm c}_{,q} )\, {\delta_{ij}}$. This is only strictly valid for a uniform initial overdensity but any departure from it would leave no imprint on angular fluctuations over the surface of the shell. Then, 
taking into account that the potential perturbations are related to the local initial density contrast through  Poisson's equation
$D' \mathrm{Tr}(\psi_{,ij}^{\rm l}) = D(a) \delta^{\rm l,i}$, we can write
\begin{equation}
\label{eq:somelabel}
D' \, \mathrm{Tr}(  \Gamma_{ij}^{-1}  \psi_{,ij}^{\rm l} ) \simeq D(a) \delta^{\rm l,i} \left( 1+D '  \frac{1}{q}\psi^{\rm c}_{,q} \right) \;.
\end{equation}
In Eqs.~(\ref{eq:zeldo1}-\ref{eq:zeldo2}), $1+\delta$ refers to the contrast with respect to the background mean density. However, we are interested in the contrast of fluctuations in the shell that are in excess of the smooth cluster contribution $1+\delta^{\rm c}$. We can therefore multiply Eq.~\eqref{eq:zeldo2} by $\vert \Gamma \vert$. We also simplify terms involving the derivatives of the potential by considering the small initial cluster density contrast $ \delta^{\rm c,i}$  (at radius $q$) and the mean initial density contrast  $ \bar{\delta}^{\rm c,i}$ (averaged inside the sphere of radius $q$) 
\begin{equation}
D' \frac{\psi^{\rm c}_{,q}}{q}  =  \frac{D(a)\,\bar{\delta}^{\rm c,i}}{3}  \equiv \frac{D(a)}{q^3} \int_0^q \delta^{\rm c,i}  p^2 \der p \;.
\end{equation}
When expressed relative to the smooth cluster density, Eq~\eqref{eq:zeldo2} becomes
\begin{equation}\label{eq:zeldo2b}
\frac{ 1+ \delta }{1+\delta^c} = \vert \Gamma \vert \, ( 1+ \delta )  \simeq  1 +  D(a) \delta^{\rm l,i} \left[ 1 + D(a) \frac{\bar{\delta}^{\rm c,i}}{3}  \right]\;,
\end{equation}
noticing that without the cluster one would recover the classical linear theory result $\delta = D(a) \delta^{\rm l,i}$. Local fluctuations experience a multiplicative boost factor corresponding to the term in brackets
in Eq.~\eqref{eq:zeldo2b}, and thus the power spectrum of local fluctuations in the cluster vicinity can be written as 
\begin{equation} \label{eq:zeldo3}
P_{\rm cluster} (\vec{k})  = P_{\rm random}(\vec{k}) \left[ 1 +  \frac{D(a) \, \bar{\delta}^{\rm c,i}}{3}\right]^2  \equiv  \alpha\; P_{\rm random}(\vec{k}) \;.
\end{equation}
In this equation, $P_{\rm random} = D(a)^2 P_0$, as expected from  linear theory. 
We nevertheless assume below that the relation still holds for the  nonlinearly evolved  $P_{\rm random}$.

Eq.~\eqref{eq:zeldo3} simply quantifies the boost imposed by  the peak condition in the initial conditions $\bar{\delta}^{\rm c,i}$. This is  consistent with the perturbative approach of \citet{Codis++17} who showed that  gravitational  evolution  induces a nonlinear bias at all multipoles proportional to the peak height $\nu$, the amplitude of fluctuations $\sigma_{0}\propto D(a)$ and the rescaled three-point function $\xi^{(3)}$. The agreement between both approaches follows from the $D(a)\bar{\delta}^{\rm c,i} \leftrightarrow \sigma_{0}\nu$ correspondence. 
From Eq.~\eqref{eq:zeldo3} and \citet{Codis++17}, it appears that the excess amplitude of harmonics  scales like $D(a)$ at first order. 
However, these predictions are only valid in the weakly nonlinear regime.

\subsection{The highly nonlinear regime}
The above formalism  can be extended to  a fully  nonlinear regime,  where gravitational clustering 
boosts all multipoles in proportion, through the convergence of the flow towards the central peak.
In practice, one needs to relate $P_{\rm cluster} (\vec{k}) $ and $P_{\rm random}(\vec{k})$ to the cluster environment at late time. 
In the Zeldovich approximation (or in the spherical collapse model before shell crossing),  the Lagrangian radius $q$ is  related to its evolved Eulerian counterpart $r$  via   
\begin{equation}
 \vec{r} =  \vec{q}\, \rho(<r)^{-1/3},
 %\left[ 1 - \frac{D(a)\bar{\delta}^{\rm c,i}(q)}{3} \right] ,
 \end{equation}
if the mass enclosed by this falling shell is constant $M(<q) = M(<r)  $. Writing $q=\lambda r$, where $\lambda=\rho(<r)^{1/3}$ is related to the mean density within the sphere of radius $r$, the solution for $\lambda$  is 
\begin{align}
\lambda &=  1 +  \frac{D(a) \,\bar{\delta}^{\rm c,i}}{3}, &\quad & {\rm at\ early\ time,}  \label{eq:lambdaearly}\\ 
              &=   \left( \frac{3 M(<r) }{4\pi \bar{\rho} r^3} \right)^{1/3},  &\quad &{\rm at\ late\ time.}\label{eq:lambdalate}
\end{align}
The boost of power $\alpha$ previously defined in Eq.~\eqref{eq:zeldo3} was found to  be $\alpha =\lambda^2$ at early time and 
we assume here it should remain $\alpha =\lambda^2$ at late time.
At late time, it is also convenient to assume a NFW density profile \citep{NFW97} for the equilibrium state of the clusters, which is characterised by is virial mass $M_{\rm vir}$ and concentration $c$. These two parameters are correlated and slowly change with time. This allows us to express the density enclosed in the sphere of radius $r$ as
\begin{equation}
\label{eq:NFW}
\rho(<r)=\rho_{\rm vir} \frac{f_{\rm NFW}( c r/R_{\rm vir})}{ f_{\rm NFW}(c)}\,,
\end{equation}
with $\rho_{\rm vir}$ the mean density inside the virial radius $R_{\rm cir}$ and $f_{\rm NFW}(x)  = [\log(1+x)- x /(1+x)] / x^3$, which relates the mean density contrast at radius $r$ to the contrast at the virial radius.  We finally get
\begin{equation}\label{eq:lambdalateNFW}
\alpha =   \left[\frac{ \rho_{\rm vir}}{\bar \rho} \frac{f_{\rm NFW}( c r/R_{\rm vir})}{ f_{\rm NFW}(c)} \right]^{2/3}\;.
\end{equation}
where $\bar \rho$ is the mean background density.
Eq.~\eqref{eq:lambdalateNFW} only depends on time via the (weak) time and mass dependence variation of the concentration parameter \citep{Klypin16}. This can be see in Fig.~\ref{fig:alpha_z}.

%------------------------------------------------------------------------------ 
%------------------------------------------------------------------------------ 
\section{Spectrum of multipolar moments at random location}\label{app:checkrandom}

%.................
\begin{figure}
\centering
\includegraphics[width=0.5\textwidth]{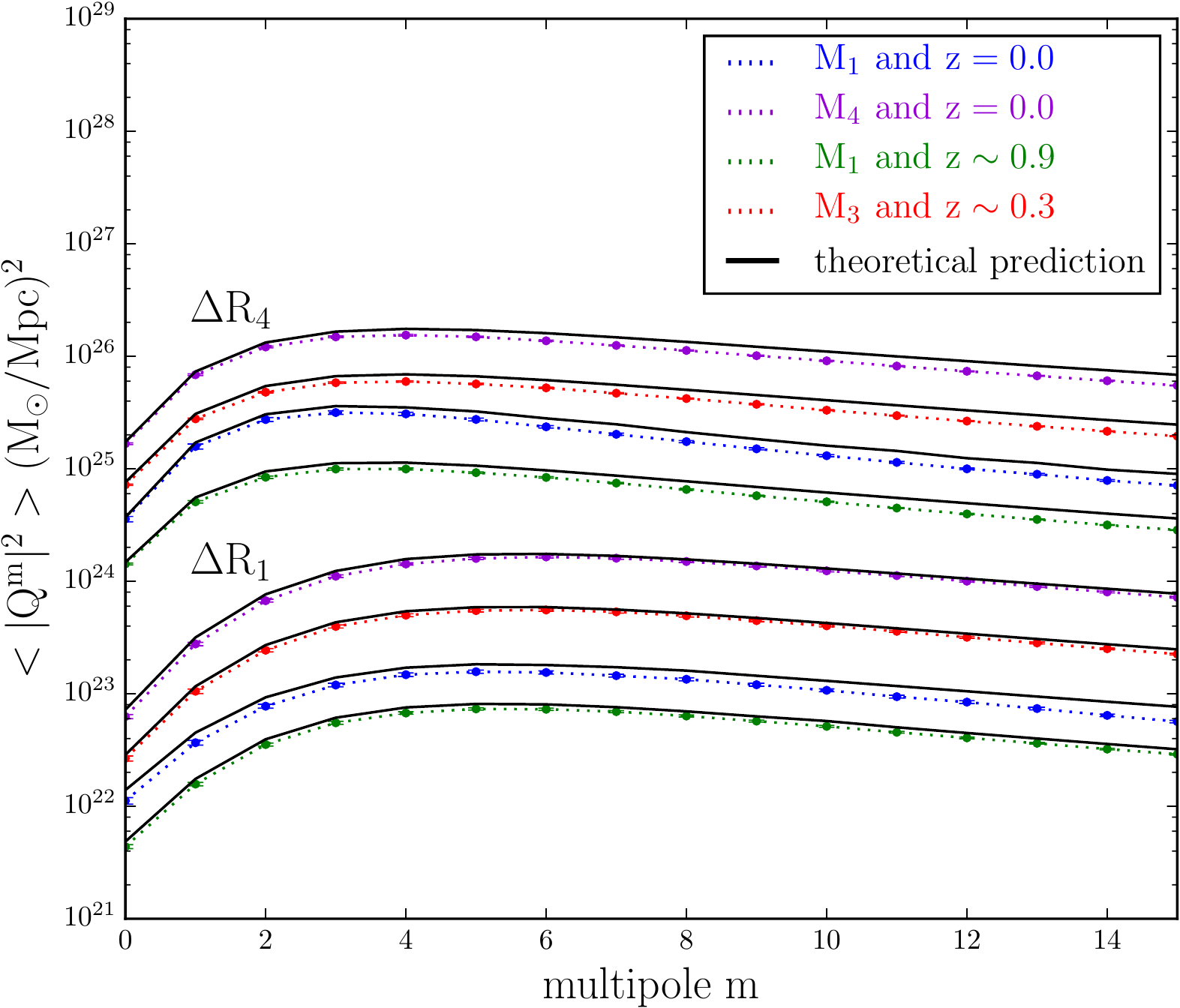}
\caption{Comparison between the analytical (solid lines) and the measured (dotted lines with tiny error bars) spectrum of multipolar moments for annuli taken at random positions in the simulation box. The bottom (resp. top) curves correspond to small $R \in [0.25-0.5]\,R_{\rm vir}$  (resp. large $R \in [1-4]\,R_{\rm vir}$) annuli. Four mass and redshift values are overlaid for each radius.}
\label{fig:comp_meas_theo}
\end{figure}
The spectrum of multipolar moments calculated at random locations can be inferred from the power spectrum of density fluctuations and compared to the spectrum measured in simulations. In Sect.~\ref{ssec:anal2ptRL}, we related the spectrum of multipolar moments with the power spectrum of the underlying two-dimensional density (or, there, convergence) field. We thus need to compute the power spectrum of the projected density $\Sigma(\vec{R})$ from the power spectrum of the three-dimensional density   fluctuations $P_\delta(\vec{k})$. 
Since we excise spheres of size $R_{\rm H}= 4 R_{\rm vir}$, the statistical properties of the projected density from this sphere are not stationary because the radial extent $a$ over which the density is integrated is a function of the projected radius $R$,  $a=a(R)=\sqrt{R_{\rm H}^2-R^2} $. Therefore,    the two-dimensional spectrum reads
\begin{equation}
P_{\Sigma} (k_{\perp}, R)  =  \frac{2 a^2 \ \overline{\rho}^2}{\pi}  \int^{\infty}_{-\infty} \der k_\parallel  \,\ P_{\delta} \left(\sqrt{k_{\perp}^2+k_\parallel^2}\right) \, {\rm sinc}\left(k_\parallel  a \right)^2 \;.
\end{equation}
Fig.~\ref{fig:comp_meas_theo} compares this analytical prediction with the measurements made in the simulation for different masses, annuli and  redshifts. For this calculation, we again use the Boltzmann code {\tt CLASS} toolkit \citep{class,class2} for the fiducial Planck Cosmology. To simplify the expression of the theoretical prediction, we approximate $a(R) =\sqrt{R_{\rm H}^2-\bar{R}^2} $ with $\bar{R} = \sqrt{R_{\rm min}\,R_{\rm max}}$, the geometrical mean radius of the annulus. 
The agreement is quite satisfactory. The small mismatch is due to the simplifying assumptions used to conduct the analytical integration (approximation on $a(R)$), and to the theoretical uncertainties on the nonlinear power spectrum $P_\delta$ on the smallest scales.
%.................

%------------------------------------------------------------------------------ 
%------------------------------------------------------------------------------ 
\section{Covariance of multipolar moments}\label{app:covar}
Let us present here the details of the derivation of the four-point statistical properties of the shape noise contribution to the local multipolar moments in Eq.~\eqref{eq:covariance}, by following the same notations as \citet{schneider98}.  Let us neglect terms that are due to the finite number of sources carrying shear. No coherent shear   is carried by galaxies either. Hence, uncorrelated galaxy ellipticities satisfy
\begin{equation}
 \langle \epsilon_{i\alpha} \rangle = 0\,,  \quad
 \langle \epsilon_{i\alpha} \epsilon_{j\beta} \rangle =  \sigma_\epsilon^2 \delta_{ij}\delta_{\alpha\beta}\,,
\end{equation}
with Latin indices $i\in {1\ldots N}$ labelling different galaxies in an annulus $[\nu \rmax, \rmax]$ and Greek indices labelling ellipticity components $\alpha \in \{ t,\times\}$, hence following the notations of \citet{Schneider02} with the distinction that $\sigma_\epsilon$ is, here, the one-dimensional dispersion of source ellipticities. 
The four-point expectation value of ellipticities is
\begin{equation}\label{eq:4ptell}
\langle \epsilon_{i\alpha} \epsilon_{j\beta}  \epsilon_{k\mu} \epsilon_{l\nu} \rangle = 
      \sigma_\epsilon^4 \left(  \delta_{ij}\delta_{\alpha\beta}\delta_{kl}\delta_{\mu\nu}  + \delta_{ik}\delta_{\alpha\mu}\delta_{jl}\delta_{\beta\nu}  + \delta_{il}\delta_{\alpha\nu}\delta_{jk}\delta_{\beta\mu}  
               \right) \nonumber\,.
\end{equation}
This will be useful for averaging over source ellipticities the multipolar moment power spectrum
\begin{equation}\label{eq:qm2estim}
\begin{split}
\vert Q_m \vert^2 &= &\frac{1}{\overline{n}^2}\sum_{jk} {\rm e}^{\imath  m ( \varphi_j - \varphi_k)}  \Big[\left(
     \beta^m_{jt}\beta^m_{kt} \epsilon_{jt}\epsilon_{kt} + \beta^m_{j\times}\beta^m_{k\times} \epsilon_{j\times}\epsilon_{k\times} 
 \right) \\
 & & + \imath \left( 
   \beta^m_{j\times}\beta^m_{kt} \epsilon_{j\times}\epsilon_{kt} - \beta^m_{jt}\beta^m_{k\times} \epsilon_{jt}\epsilon_{k\times} 
 \right) \Big],
\end{split}
\end{equation}
 adopting the convention 
%\begin{equation}\label{eq:bdef2}
$\beta^m_{j\alpha} = b_{\alpha,m}(r_j)/r_j$. 
%\end{equation}
In the absence of shear, the expectation value of \eqref{eq:qm2estim} is simply
\begin{equation}\label{eq:shapenoisemean2ptDISC}
\langle \vert Q_m \vert^2 \rangle \equiv  \vert Q_m^{{\rm s}} \vert^2 = \frac{\sigma^2_{\epsilon} }{\overline{n}^2 }  \sum_{k=1}^N  \left( \beta^{m,2}_{kt} + \beta^{m,2}_{k\times}\right) \;.
\end{equation}
Accounting for symmetries and arranging terms, the four-point moments reads
\begin{equation}\label{eq:shapenoisemean4ptALL}
\langle \vert Q_m \vert^2 \, \vert Q_n \vert^2 \rangle =  
   \langle \vert Q_m \vert^2 \rangle  \, \langle \vert Q_n \vert^2  \rangle + \vert K^{+}_{mn}\vert^2  + \vert K^{-}_{mn}\vert^2  \,,
\end{equation}
where 
\begin{equation}\label{eq:AmnGmn}
K^\pm_{mn} \equiv \frac{\sigma^2_{\epsilon} }{\overline{n}^2 } \sum_k {\rm e}^{\imath  (m \mp n) \varphi_k} \left(  \beta^m_{kt} \beta^n_{kt} \pm\beta^m_{k\times} \beta^n_{k\times} \right) \;.
\end{equation}
We highlight that the averaging over source positions within the annulus yields 
\begin{equation}\label{eq:Amn}
\langle K^+_{mn} \rangle = \delta_{mn} \vert Q_m^{\rm s} \vert^2 \,,\quad
\langle K^-_{mn} \rangle = \delta_{m0} \delta_{n0} \tilde{Q}_0^{{\rm s}2} ,
\end{equation}
with the modified moments defined by
\begin{eqnarray}\label{eq:shapenoisemean2ptDISCalt}
\vert Q_m^{\rm s} \vert^2  &= &  \frac{\pi \sigma^2_{\epsilon}}{\overline{n} } \int_{\nu R}^R dr \frac{b_{t,m}^2(r)+b_{\times,m}^2(r)}{r}\,,\\
\vert \tilde{Q}_m^{{\rm s}} \vert^2 &= & \frac{\pi \sigma^2_{\epsilon}}{\overline{n} } \int_{\nu R}^R dr \frac{b_{t,m}^2(r)-b_{\times,m}^2(r)}{r}\,.
\end{eqnarray}

\end{document}